\documentclass[aps,pra,floatfix,reprint, twocolumn, nofootinbib,notitlepage]{revtex4-1}
\usepackage[caption=false, subrefformat=parens, labelformat=parens]{subfig}
\usepackage{graphicx}
\usepackage{dcolumn}
\usepackage{bm}
\usepackage{amsmath, amsthm}
\usepackage{rotating}
\usepackage{amsmath,amsthm,amssymb,mathtools}
\usepackage{algorithm}
\usepackage[noend]{algpseudocode}
\usepackage{verbatim}
\usepackage{float}
\usepackage{outline}
\usepackage{tikz}
\usepackage{empheq}
\usetikzlibrary{shapes.geometric, arrows}
\usepackage{braket}
\usepackage{colonequals}
\usepackage{textcomp}
\usepackage[inline]{enumitem}
\usepackage[colorlinks]{hyperref}
\usepackage[all]{hypcap}
\usepackage[utf8]{inputenc}
\usepackage[english]{babel}
\usepackage{csquotes}

\newtheorem{lemma}{Lemma}

\theoremstyle{definition}

\theoremstyle{remark}

\newtheorem{problem}{Problem}

\def\QAOA{\mathrm{QAOA}}

\newcommand{\z}{\sigma^z}
\newcommand{\x}{\sigma^x}
\newcommand{\y}{\sigma^y}
\newcommand{\Tr}{\text Tr}
\def\ket#1{|#1\rangle}

\def\ketbra#1#2{|#1\rangle\langle #2|}
\newcommand{\bqa}{\begin{eqnarray}}
\newcommand{\eqa}{\end{eqnarray}}
\newcommand{\beq}{\begin{equation}}
\newcommand{\eeq}{\end{equation}}

\newcommand{\eq}{&=&}

\newcommand*\widefbox[1]{\fbox{\hspace{2em}#1\hspace{2em}}}

\newcommand{\nnm}{\nonumber\\}

\newcommand{\zw}[1]{{\color{blue}  #1}}

\newcommand{\egr}[1]{{\color{black}  #1}}

\newcommand{\ZtotV}{Z_{tot,v}}

\newcommand{\phaseHam}{H_{\mathrm{PS}}}
\newcommand{\penHam}{H_{\text{pen}}}
\newcommand{\penFun}{f_\text{pen}}
\newcommand{\identity}{\mathbf 1}

\newcommand{\HPS}{H_{\text{PS}}}
\begin{document}
\title{XY-mixers: analytical and numerical results for QAOA}
\author{Zhihui Wang}
\email{zhihui.wang@nasa.gov}
\affiliation{Quantum Artificial Intelligence Laboratory (QuAIL),
NASA Ames Research Center, Moffett Field, CA 94035}
\affiliation{Universities Space Research Association, 
615 National Ave, Mountain View, CA 94043 }
\author{Nicholas C. Rubin}
\email{nickrubin@google.com}
\affiliation{Google Inc., 340 Main Street, Venice, CA 90291, USA}
\affiliation{Rigetti Quantum Computing, 775 Heinz Ave, Berkeley, CA 94710}
\author{Jason M. Dominy}
\affiliation{Department of Applied Mathematics, University of California, Santa Cruz, Santa Cruz, CA 95064}
\author{Eleanor G. Rieffel}
\affiliation{Quantum Artificial Intelligence Laboratory (QuAIL),
NASA Ames Research Center, Moffett Field, CA 94035}

\begin{abstract}
The Quantum Alternating Operator Ansatz (QAOA) is a promising gate-model meta-heuristic for combinatorial optimization. 
Applying the algorithm to problems with constraints presents an implementation challenge for near-term quantum resources. 
This work explores strategies for enforcing hard constraints by using $XY$-Hamiltonians as mixing operators (mixers).  Despite the complexity of simulating the $XY$ model, we demonstrate that for an integer variable admitting $\kappa$ discrete values represented through one-hot-encoding, certain classes of the mixer Hamiltonian can be implemented without Trotter error in depth $O(\kappa)$.  We also specify general strategies for implementing QAOA circuits on all-to-all connected hardware graphs and linearly connected hardware graphs inspired by fermionic simulation techniques. Performance is validated on graph coloring problems that are known to be challenging for a given classical algorithm.  The general strategy of using $XY$-mixers is borne out numerically, demonstrating a significant improvement over the general $X$-mixer, and moreover the generalized $W$-state yields better performance than easier-to-generate classical initial states when $XY$ mixers are used.
\end{abstract}
\maketitle
\section{Introduction}
Prior to achieving full error-correction, which likely requires large physical-to-logical qubit ratios and low error-rates, the exploration of what near-term quantum resources can achieve is paramount. \egr{One of the main uses of near-term quantum devices will be to evaluate quantum algorithms beyond the reach of classical simulation.} One of the most exciting and anticipated \egr{potential} uses of quantum computers is solving combinatorial optimization problems, \egr{with near-term quantum hardware providing unprecedented means for exploring and evaluating quantum algorithms for optimization}.  The quantum-approximation-optimization algorithm (QAOA) has risen \egr{to be the leading} candidate to test the applicability of gate-model quantum resources at solving optimization problems \egr{on near-term quantum hardware} prior to fault-tolerance~\cite{farhi2014quantum, hadfield2017quantum, farhi2017quantum}.  Studies using QAOA to obtain the $\Theta(\sqrt{2^n})$ query complexity on Grover's problem,\cite{Jiang17}, to find approximate solutions to MAXCUT~\cite{farhi2014quantum, wang2017quantum, zhou2018quantum}, MAXE3LIN2~\cite{farhi2014quantum_bounded_occurrence}, network detection~\cite{shaydulin2018network}, simple machine learning models~\cite{farhi2018classification, otterbach2017unsupervised} and sampling~\cite{Farhi2016} suggest that there is a path forward to obtaining high quality solutions with QAOA under a noiseless environment.   The hybrid nature of this algorithm implies that noise of physical qubits can be tolerated to some extent~\cite{PhysRevX.6.031007, farhi2017quantum, zhou2018quantum}.

These initial findings led to the development of a general framework known as the Quantum Alternating Operator Ansatz (also QAOA) that extends the utility of the initial algorithm to a wide variety of optimization problems involving linear or non-linear constraints \egr{and to a wider variety of mixing operators that can greatly increase the implementability of a QAOA approach to many combinatorial optimization problems}~\cite{hadfield2017quantum}.  Both frameworks are \egr{meta-heuristics, so}
require \egr{further} specification. \egr{Challenges include devising}  strategies for selecting angles with minimal computational overhead, efficient initial state determination and preparation, and embedding high-dimensional graphs--e.g. non-planar graphs--into physically realizable lattices of qubits.  
\egr{Prior work on} components of the general QAOA algorithm for handling hard and soft optimization constraints include lattice protein folding by changing the driver~\cite{fingerhuth2018quantum}, classical and quantum embeddings for representing all-to-all connected graph problems~\cite{lechner2018quantum}, 
optimization strategies~\cite{zhou2018quantum}, and compilation strategies~\cite{guerreschi2018qaoa, crooks2018performance,Venturelli18_Compiling, Booth18, PhysRevX.7.021027} 
further our understanding how to apply the QAOA heuristic.

In this work we explore the feasibility of expanding QAOA's scope of applicability to discrete optimization problems with integer variables, as conceptually proposed in Ref.~\onlinecite{hadfield2017quantum}.  Commonly, $\kappa$-ary variables facilitate simpler representations of combinatorial optimization problems and open the possibility of multiple encoding strategies.  For example, integer variables can be directly encoded into binary, redundantly encoded in a classical coding fashion, or into a one-hot-encoded set if $\kappa$ is small.  In this work we study the implementation and performance of one-hot-encodings for graph coloring problems with QAOA using mixers based on $XY$-Hamiltonian.  Pairing the one-hot-encoding and $XY$-mixers is a natural choice as $XY$-mixers preserve the representation~\cite{hadfield2017quantum}.  To validate the $XY$-mixing Hamiltonian is consistent with the salient feature of QAOA--short depth circuits--we provide short depth circuit implementations for each term in QAOA.  
For any one-hot-encoded integer variable problem, the feasible subspace is spanned by all Hamming-weight-1 bit strings.  Given such a problem on an all-to-all connected hardware platform, we propose a scheme that can generate the exact evolution of the $XY$-model on a complete graph in linear depth.
Moreover, exploiting the fermionic transformation, we show that the $XY$ model on a ring can be realized in logarithmic depth. 
Most notably, due to the commuting nature of the cost Hamiltonians~\cite{hadfield2018representation} a SWAP-network, akin to sorting networks, can be used to implement any $2$-local cost operator requiring all-to-all connectivity in linear depth on a linearly connected graph of qubits.  Though the $XY$-mixer is significantly more complicated than the standard $X$-mixer, we demonstrate that under numerous scenarios this driver term can be implemented in linear depth by taking a fermionic perspective.  If approximate evolution is found to be tolerable, for all-to-all connected architectures, the first-order Trotter implementation of the $XY$-mixer drops to $O(\log(\kappa))$ circuit depth.  

Through numerical simulations, we also compare performance of different $XY$ mixers. 
In a noise-free scenario, the mixer based on $XY$ model on a complete graph $K_\kappa$ for each node gives better performance than the mixer using the $XY$ model on a ring for finite QAOA levels.  This advantage needs to be considered as a tradeoff to the complexity of the circuit generating the mixing unitary; furthermore, the realistic performance will also depend on the effects of noise and gate infidelity.
We also show that initial states play a crucial rule in QAOA with $XY$ mixers.  While an easy-to-generate classical state serves as a valid initial state in the feasible subspace, the generalized W-state, i.e., the uniform superposition of all Hamming-weight-one bit strings, yields significantly better performance.

The rest of the paper is structured as follows:
Section~\ref{sec:QAOA_framework} outlines the general QAOA framework, 
with the emphasis on an analysis showing the approximation ratio for optimization problems in a discrete bounded domain can provide a lower bound on the typical case,
Section~\ref{sec:problem_formulation} formulates the 
Max-$\kappa$-Colorable-Subgraph problem in a binary representation and introduces the terminology required for comparing the $XY$- and $X$-mixers,
Section~\ref{sec:circuit_XY} describes methods for implementing various mixers in short-depth circuits. 
In Section~\ref{sec:triangle}, $XY$ mixer is demonstrated to outperform the $X$-mixer and in~\ref{sec:small_and_hard} and~\ref{sec:benchmark} we provide benchmark numerics on small hard-to-color graphs, and all $\kappa$-colorable graphs of given sizes.
Circuit implementation strategies for 
$W$-state generation
are relegated to the Appendix. 
\section{The QAOA framework}\label{sec:QAOA_framework}
The QAOA framework starts with the specification of a cost Hamiltonian (phase-separating Hamiltonian) $\HPS$ such that its specification requires a polynomial number of $k$-local terms that all commute.  Commonly, the $z$-computational basis states are used for problem encoding.  That way, every basis state corresponds to an eigenstate of $\HPS$.  The objective is to to find the lowest energy eigenstate by a quantum evolution that effects transition between the eigenstates.  The $\HPS$ term serves to interfere various eigenstates and thus change the transition probabilities.

A QAOA circuit of level $p$ consists of the following steps:
\begin{enumerate}
\item Prepare a suitable initial state $\ket{\psi_0}$;
\item Repeat the following steps $p$ times: in the $l$-th repetition,
apply the phase-separating unitary $\exp[-i\gamma_l H_{PS}]$
and
apply the mixing unitary $U_{M}(\beta_l)$;
\item Measure in the computational basis.
\end{enumerate}
The unitaries are parametrized by a set of real numbers 
$\{\gamma_j,\beta_j\}_{j=1}^p$, respectively. In a classical-quantum hybrid setup, Monte Carlo averaging for the expectation value of $\HPS$ serves as the objective function for classical feedback on the angles $\{\gamma_j,\beta_j\}_{j=1}^p$. Efficient strategies for statistical estimation of $\langle \psi(\mathbf{\gamma}, \mathbf{\beta}) \HPS \psi(\mathbf{\gamma}, \mathbf{\beta})\rangle$ and for noncommuting Hamiltonians have been discussed in References~\cite{mcclean2016theory, rubin2018application}.

In many QAOA case studies, analytical or numerical,
the expectation value of $\HPS$ instead of the probability of the lowest energy solution 
has been used as a proxy for performance.
Concern has been raised in using expectation values (or approximation ratio) as a sole figure of merit
because a high expected value does not guarantee the quality of solutions upon measurement, 
the underlying distribution needs to be scrutinized.  In a general setting, the variance of the distribution would be required to further indicate sufficient concentration of probability on the desired solution. 

We point out that for problems with a domain of discrete real values, as most combinatorial optimization problems are,
a high approximation ratio generally accompanies a high value for the typical case.
To see this, we analyze how the tail probability is bounded by the mean 
when the domain is a set of bounded discrete integer values.
Consider a random integer variable $X \in \{0,1, \dots,m\}$; 
if the mean value is $\mu$ 
then for any $l \le \lfloor \mu \rfloor$,
where $\lfloor \cdot \rfloor$ is the floor function,
the probability of $x$ taking a value larger than $l$ is lower-bounded as
\bqa
\Pr(X > l) \ge  \frac{\mu -l} {m- l}\;.
\label{eq:bound}
\eqa
In Appendix.~\ref{app:bound} we provide a proof for Eq.~\eqref{eq:bound} under more general assumptions.
In Sec.~\ref{sec:benchmark} we will see examples: for our QAOA results of high approximation ratio, without examining the energy distribution,
we can infer with high confidence that a typical solution will have high cost.

\section{Problem Formulation}
\label{sec:problem_formulation}
In this section we formulate the Max-$\kappa$-Colorable-Subgraph problem in binary form using a one-hot-encoding representation for the possible colorings of each node.  Using binary variable $x_{v,c}$ to indicate whether vertex $v$ is assigned color $c$, the one-hot-encoding formulation requires solutions live in a subspace of the full Hilbert space that satisfies: $\sum_{c}x_{v,c}=1$, the \emph{feasible subspace}.  This results in two formulations of the coloring problem for QAOA--with and without a penalty term in the phase-separator.  We also recap nomenclature for various $XY$-Hamiltonian drivers introduced in Ref.~\cite{hadfield2017quantum} which becomes necessary when discussing their circuit implementations.  

The Max-$\kappa$-Colorable-Subgraph problem is formulated as follows:
\begin{problem}
Given a graph $G = (V, E)$ with $n$ vertices and $m$ edges, and $\kappa$ colors, 
maximize the size (number of edges) of a properly vertex-colored subgraph.
\end{problem}

\begin{figure}[htbp]
    \centering
    \includegraphics[width=0.7\columnwidth]{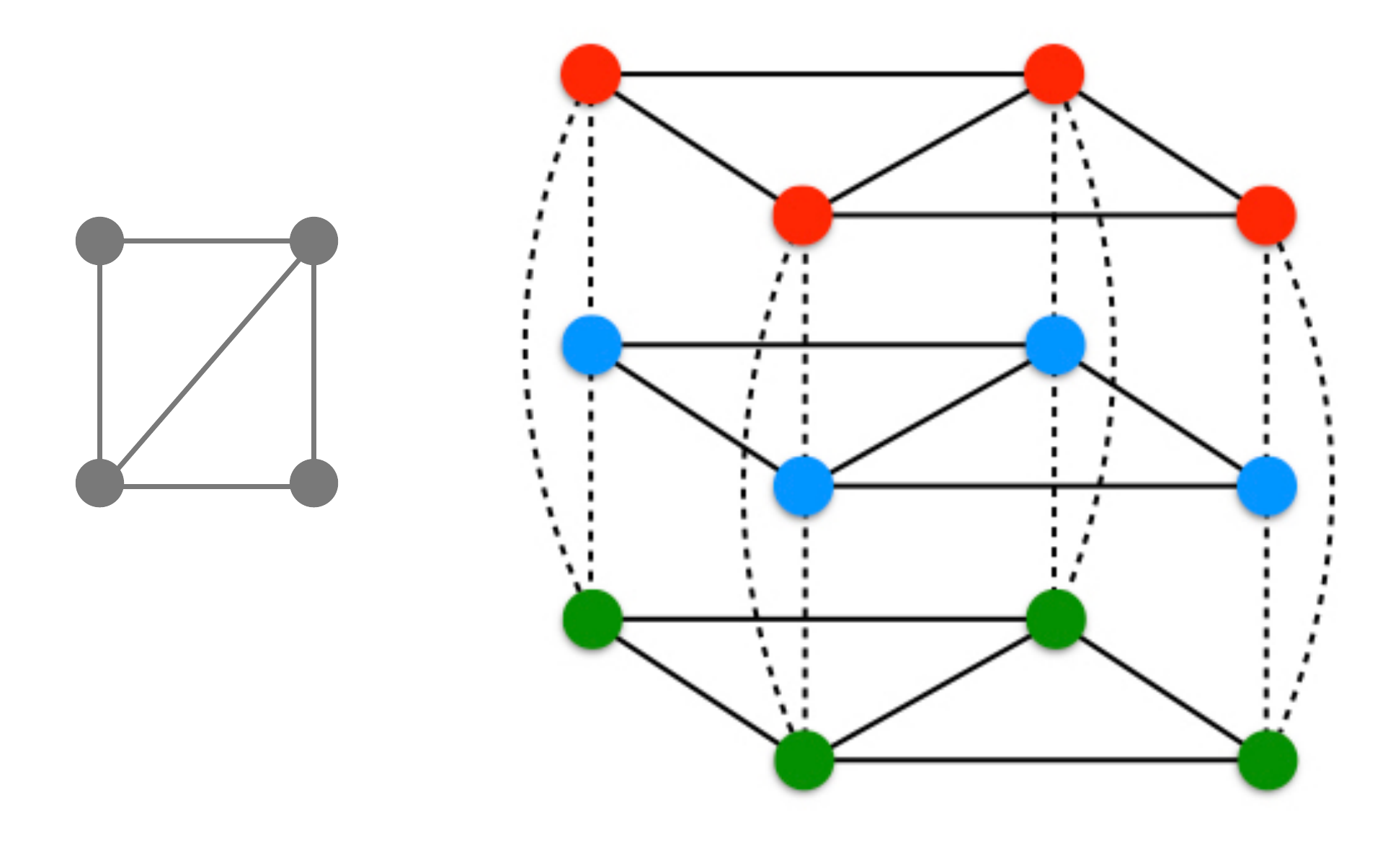}
    \caption{
    Left: The original graph to-be colored.
    Right: The qubit-layout encoding the problem.
    Each vertex $v$ is represented by $\kappa$ qubits $x_{v,c}$ for $c=1,\ldots \kappa$ representing its $\kappa$ possible colors.  The extended graph construction can be thought of taking the graph represented in its natural Euclidean space and then augmenting that space with another dimension and replicating the graph $\kappa$ times for each of the colors.  
    The phase separation Hamiltonian is composed of two-qubit operations corresponding to edges on each surface,
    and the mixing operation is among the qubits in the augmented dimension.
    }
    \label{fig:color_encoding}
\end{figure}

The max-$\kappa$-Colorable-Subgraph problem is encoded into qubits with a one-hot encoding fashion to represent the colors.
Each node of the graph $G$ is expanded into $\kappa$ qubits where each qubit occupation is used to represent a coloring of the node.  
For example, a three-coloring problem on a graph of four vertices requires 12 qubits depicted in Figure~\ref{fig:color_encoding}.  

In the feasible subspace where each vertex is assigned exactly one color, the cost/objective function
\bqa
\label{eq:fC}
f_C= m-\sum_{j=1}^\kappa \sum_{\{v,v'\}\in E} x_{v,j}x_{v',j} 
\eqa
counts the properly-colored edges,
and we aim at maximizing $f_C$.
By the replacement $x\to (\identity - \z)/2$ in Eq.~\eqref{eq:fC}, the corresponding quantum objective Hamiltonian is
\bqa
H_C 
\label{eq:HC}
\eq \frac{1}{4}\Big( (4-\kappa)m\identity + H'_C \Big)\;,
\eqa
where
\bqa
H'_C \eq  \sum_{v=1}^n d_v \sum_{j=1}^\kappa \z_{v,j} - \sum_{j=1}^\kappa \sum_{\{v,v'\}\in E} \z_{v,j}\z_{v',j}\;.
\eqa
Throughout this paper we use $\sigma_x$, $\sigma_y$, $\sigma_z$, and $X$, $Y$, $Z$ interchangeably to refer to the Pauli operators.
The approximation ratio we will adopt in the following is the ratio of the expectation value of the cost Hamiltonian, projected onto the feasible subspace, to the true maximal number of correctly colored edges:
\begin{align}
\label{eq:def_approx_ratio}
    r = \frac{\langle P_{\mathrm{feas}} H_C P_{\mathrm{feas}} \rangle}{C_{\max}},
\end{align} 
where $P_{\mathrm{feas}}$
is the projection operator onto the feasible subspace, and $C_{\max}$ is the number of edges in the true max-$\kappa$-colorable subgraph.
Note that by projecting to the feasible subspace, the ratio of the feasible subspace to the full Hilbert space is also factored in.
The numerator in Eq.~\eqref{eq:def_approx_ratio} is equivalent to the average number of properly-colored edges observed upon measurement, 
with the infeasible output valued zero.

\subsection{Adding a penalty in the phase separating Hamiltonian}
\label{sec:penalty}
The common practice for incorporate constraints is to add a penalty term to the cost function. For the one-hot-encoded problem we define a quadratic penalty to penalize the case that a node is assigned no color or multiple colors
\bqa
\penFun \eq \sum_v (1-\sum_{j=1}^\kappa x_{v,j})^2 
\eqa
which, up to a constant, corresponds to the penalty Hamiltonian, 
\bqa
\penHam 
\eq \frac{1}{2} \sum_v \Big((2-\kappa) \sum_{j}  \z_{v,j} + \sum_{j<j'} \z_{v,j}\z_{v,j'} \Big)
\eqa
that increases the energy of all states outside the subspace. The phase-separating Hamiltonian becomes a weighted sum of the cost and the penalty Hamiltonians
\bqa
\label{eq:phaseHam}
\phaseHam \eq H'_C - \alpha \penHam
\eqa
where the weight parameter $\alpha \in \mathcal R_+$. Note that in Eq.~\eqref{eq:phaseHam} the penalty Hamiltonian is subtracted because we aim to maximize the original cost function and minimize the penalty.  In order for the penalized function to have the same optima as the original cost function, the penalty weight needs to be set above a critical value. In the current problem, assigning more than one color to a vertex is not energetically favorable,  so it is the opposite, assigning no color to a vertex that may create fake maxima.  
Since for every no-color vertex, there are at most $\lfloor d_v/\kappa\rfloor$ edges lifted from being improper, the penalty should satisfy $\alpha > \lfloor\max\{d_v\}/\kappa\rfloor$, we can loosely take $\alpha \ge n/\kappa$.  On the other hand, the range of possible values of $f_C$ (and of spectral values of $H_C$) is $\kappa m$.  Therefore, any $\alpha > \kappa m$ will ensure an energy separation between all feasible states and all infeasible states.

It should be noted that, unlike the motivating situation in adiabatic computation, the energy gap plays no clear role in QAOA.  Thus it should be expected that, while the introduction of a penalty into the cost Hamiltonian may alter the QAOA dynamics, perhaps manipulating the reachable set of unitary operators, the role of the penalty strength is unclear at best.  This perspective is supported by the numerical results in Section \ref{sec:triangle}.  Indeed, while for some problems, such as the one-hot-encoded problems under consideration, sophisticated mixers can be designed to satisfy the constraints \cite{hadfield2017quantum}, the design of general and systematic methods for incorporating constraints into QAOA remains an open problem.

In the penalty formulation the mixer can be either the standard $X$-driver
\bqa
H_X=\sum_{v=1}^n\sum_{c=1}^\kappa \x_{v,c}\;
\eqa
or the $XY$-Hamiltonian.  If the $XY$-Hamiltonian is selected the penalty parameter may help the variational optimizer maintain probability mass in the feasible subspace and is not strictly necessary.  In QAOA, it is unclear how a penalty parameter helps maintain probability mass over the feasible subspace.  The feasible space of a $\kappa$-coloring problem is the set of states that satisfy

\begin{align}
\ZtotV \equiv \sum_{c=1}^\kappa \z_{v,c} = \kappa-2\;,
\end{align}
i.e., a subspace spanned by states in the computational basis that correspond to bit strings of Hamming weight equal to one.

Although formulating the penalty Hamiltonian facilitates the use of the standard $X$-mixer in QAOA, which can be implemented in constant circuit depth, we emphasize that the relative size of the feasible space becomes exponentially small as the graph size grows and thus a penalty formulation is sub-optimal. To see this, consider the size of the feasible subspace ${\cal H}_{\text{fea}}$;
for each node, the feasible subspace can be spanned by states corresponding to $\kappa$ Hamming-weight one bit-strings, 
hence is of dimension $k$, and the feasible subspace for the whole problem is of dimension $k^n$.
The ratio of the feasible subspace sizes to the size of the full Hilbert space is 
\begin{align}
\frac{\mathrm{dim}(\mathcal{H}_{{\text fea}})}{\mathrm{dim}(\mathcal{H})} 
= \frac{\kappa^{n}}{2^{n\kappa}} = \big(\frac{\kappa}{2^\kappa}\big)^n,
\end{align}
which for any $\kappa\ge1$ shrinks exponentially with the graph size $n$. 

\subsection{\texorpdfstring{$XY$}{XY} mixer: Enforcing evolution in the feasible subspace}
\label{sec:XY}
The $\ZtotV$ constraint can be incorporated in a natural way by selecting a mixing term that preserves the feasible subspace.   Here we use the $XY$-Hamiltonian
\bqa
\label{eq:XYv}
& & H_{XY,v} = \frac{1}{2}\sum_{c,c'\in K}^\kappa H_{XY,v,c,c'}  \\
& & H_{XY,v,c,c'} = \x_{v,c}\x_{v,c'}+\y_{v,c}\y_{v,c'}\;.
\eqa
which drives rotations in the $\{(0, 1), (1, 0)\}$ subspace of each color labeling. In the above equation the mixer applies to any color pair $c,~c'$ in a set $K$.  It can be verified that for any $K$, $[H_{XY,v}, \ZtotV]=0$.
\subsubsection{Complete vs ring mixing Hamiltonians}
In Eq.~\eqref{eq:XYv}, when the mixing-set $K$ includes all pairs, the mixer is termed a \emph{complete-graph} mixer.
An alternative is the \emph{ring} mixer in which $K$ takes a one-dimensional (1D) structure: $c'=c+1$, and we apply periodic boundary condition. 
In the same fashion, there are a variety of derivative mixers based on the $XY$-Hamiltonian, depending on the underlying connectivity between colors.
We focus on the complete-graph and ring mixers.

\subsubsection{Simultaneous vs partitioned mixers}
For a given mixing Hamiltonian,~Eq.~\eqref{eq:XYv}, 
for each node, a \emph{simultaneous} mixer exactly applies the unitary $\exp[{-i\beta H_{XY,v}}]$
while a \emph{partitioned} mixer applies the product of $\exp[{-i\beta H_{XY,v,c,c'}]}$ 
in some order of $\{(c,c')\}$.  We define the \emph{parity-partition} mixer such that a local $XY$-Hamiltonian is applied on even pairs first and odd pairs next.

The parity-partitioned mixing unitary is a first-order approximation of the simultaneous mixing unitary. Employing the Zassenhaus formula through second order
\bqa
e^{it(H_{even}+H_{odd})} \approx e^{itH_{even}}e^{itH_{odd}}e^{\frac{t^2}{2}[H_{even},H_{odd}]}
\eqa
allows us to characterize the leading error term $e^{-t^2/2[H_{even},H_{odd}]}$ as a function of $\kappa$. For simplicity, we consider even $\kappa$, so that $H_{even}$ and $H_{odd}$ contain $n/2$ commuting terms exactly.  The parity-partitioned mixer can be represented as two separate Hamiltonians
\bqa
H_{odd} \eq H^{(XY)}_{1,2}+H^{(XY)}_{3,4}+\ldots+H^{(XY)}_{\kappa-1,\kappa}\nonumber\\
H_{even} \eq H^{(XY)}_{2,3}+H^{(XY)}_{4,5}+\ldots+H^{(XY)}_{\kappa,1}
\label{eq:parity}
\eqa
where $H^{(XY)}_{j,j'}=X_jX_{j+1}+Y_{j}Y_{j+1}$ and each term $H^{XY}_{j,j+1}$ in $H_{even}$ commutes with all terms in $H_{odd}$ except for $H^{(XY)}_{j-1,j}$ and $H^{(XY)}_{j+1,j+2}$.  We can simply determine the term generated by the commutation
\bqa
\label{eq:commute_nn}
[ H^{(XY)}_{j-1,j}, H^{(XY)}_{j,j+1} ] = 2i(X_{j-1}Y_{j+1}-Y_{j-1}X_{j+1})Z_{j}
\eqa
to obtain the general form of the error term.  Therefore, $[H_{even},H_{odd}]$ is composed of $\kappa/2$ terms of type of Eq.~\eqref{eq:commute_nn}.
Since $||X_{j-1}Y_{j+1}Z_j||$ is of order $1$, we have $||[H_{even},H_{odd}]||\sim \kappa$ and therefore expect the difference between the simultaneous and the parity-partitioned mixing operators to be more prominent as $\kappa$ grows.

In the above analysis, the two mixing operators in general do not commute.
However, we only need to focus on their effects in the feasible subspace.
Here we provide analysis on the commuting relations in the feasible subspace for general $\kappa$.
Note that each $H^{(XY)}_{j,j'}$ operation in the feasible subspace corresponds to a $2\times 2$ permutation matrix.
Then $H_{even}$ and $H_{odd}$ can be identified with the two permutations $\pi,\sigma\in S_{\kappa}$ on $\kappa$ letters:
    \begin{subequations}
        \begin{align}
                \pi & = \big(0\;1\big)\big(2\;3\big)\cdots\big((\kappa-2)\;(\kappa-1)\big)\\
                        \sigma & = \big(1\;2\big)\big(3\;4\big)\cdots\big((\kappa-3)\;(\kappa-2)\big)\big((\kappa-1)\; 0\big).
                            \end{align}
                                \end{subequations}
                                Thinking of the letters arranged on a circle, these are the two possible permutations that consist of $\kappa$ disjoint nearest-neighbor transpositions.  It may be observed that
                                    \begin{subequations}
                                        \begin{align}
                \sigma\pi = \big(0\;2\;4\;\dots\;(\kappa-2)\big)\big(1\;3\;5\;\dots\;(\kappa-1)\big)^{-1}
                                                    \end{align}
                                                        \end{subequations}
                                                            is a product of two disjoint cyclic permutations and therefore is of order $\kappa/2$, i.e. $(\sigma\pi)^{\kappa/2}$ is the identity permutation.  Now we see that $\pi$ and $\sigma$ satisfy exactly the relations necessary to generate $D_{\kappa}$, the dihedral group of order $\kappa$ \cite{Robinson1995}, i.e., 
        \begin{align*}D_{\kappa} = \langle \sigma, \pi | \sigma^{2} = \pi^{2} =    (\sigma\pi)^{\kappa/2} = 1\rangle.\end{align*}  In particular, we may note that, while $D_{4}$ is an abelian group (isomorphic to the Klein four-group $\mathbb{Z}_{2}\times\mathbb{Z}_{2}$), all $D_{\kappa}$ for $\kappa>4$ are non-abelian.
Therefore only for $\kappa=4$, 
the simultaneous and the parity-partitioned ring mixers commute in the feasible subspace, hence are equivalent for QAOA.
\subsubsection{Feasible initial states}
The initial state in the standard QAOA framework with the $X$-mixer
is  $\ket +^{\otimes n}$, the even superposition over all bit strings,  
which is a fair starting point given no prior knowledge about the optimal solution.
This state is also the ground state of the $X$-mixer, 
and can be generated by performing a single-qubit Hadamard transform on each qubit.  

Under the new QAOA framework which accepts a constraint,
the full space spanned by all bit strings $\{0, 1\}^{n}$ is no longer a valid solution space.  
When the constraint, as in our case, dictates preserving the total magnetic quantum number,
\begin{align}
\label{eq:linearconstraint}
\ZtotV = \sum_{c=1}^{\kappa}\sigma_{c}^{z} = C
\end{align}
where $C \in [-\kappa, \kappa]$ is a constant integer,
the feasible solution space is composed of
Hamming-weight $(\kappa+C)/2$ bit strings, 
which correspond to states that satisfy Eq.~\eqref{eq:linearconstraint}.
In analogy to the $\ket +^{\otimes n}$ state for the case where all bit strings are feasible solutions,
a fair initial state should be the even superposition of all Hamming-weight $(\kappa+C)/2$ bit strings. 
Such a state is also an eigenstate of the $XY$ mixer. In the graph coloring problem, $C=\kappa-2$, the generalized $W$-state is the fair starting state.  In section~\ref{sec:W} we survey circuit construction methods that can be used to create a $W$-states.
\section{Circuit realizations} \label{sec:circuit_XY}
In this section we describe how to implement the various components of QAOA  into short depth circuits.
We start by assuming the physical qubits are all-to-all connected, and show that the simultaneous complete-graph and ring mixers
can be realized in depth linear and logarithmic in $\kappa$, respectively.
In Sec.~\ref{sec:compilation}
we discuss the depth required due to limited connectivity between the physical qubits.
\subsection{Logarithmic depth simultaneous ring mixer}
\label{sec:xy_ffft}
Interacting spin-1/2 chain is one of the oldest problems of quantum mechanics.
Stemming from the resemblance between spin-1/2 raising (resp. lowering) operators and 
fermionic creation (resp. annihilation) operators, 
in a detailed work of [\onlinecite{JordanWigner}],
the Jordan-Wigner transformation was introduced 
to convert spin-1/2 systems into problems of interacting spinless fermions.
While in general spin-spin interactions map to non-local fermionic interactions,
for the one-dimensional $XY$ problem the transformation results in a particularly precise form
involving only quadratic fermionic couplings:
\begin{align}
H_{XY} = & \sum_{c = 1}^{\kappa}\left(\sigma_{c}^{x}\sigma_{c+1}^{x} + \sigma_{c}^{y}\sigma_{c+1}^{y}\right)\nonumber \\
\downarrow & \nonumber \\
H_{XY} = & 2\sum_{c=1}^{\kappa}\left(a_{c}^{\dagger}a_{c+1} + \mathrm{h.c.}\right) \label{eq:quadratic_fermion_h}\;,
\end{align}
where $\hat a$ and $\hat a^\dag$ are fermionic operators, 
and we assumed $\kappa$ is even for simpler demonstration.

The quadratic Hamiltonian in Eq.~\eqref{eq:quadratic_fermion_h} can be diagonalized by a basis rotation on the operators.  For nearest-neighbor, one-body coupling, the fermionic Fourier transform
\begin{align}\label{operator_FT}
\hat a_c^\dag &= \mathrm{FFFT}^{\dagger}\hat{f}_{k}^{\dagger}\mathrm{FFFT} \equiv \frac{1}{\sqrt{\kappa}}\sum_{c}e^{i 2\pi ck }\hat{f}_{k}^{\dagger} \\
\hat a_c &= \mathrm{FFFT}^{\dagger}\hat{f}_{k}\mathrm{FFFT} \equiv \frac{1}{\sqrt{\kappa}}\sum_{c}e^{-i 2\pi ck}\hat{f}_{k}\;, \nonumber
\end{align}
is sufficient to diagonalize the Hamiltonian.  We use the notation $\rm FFFT$ (fermionic fast Fourier transform) 
to denote the circuit for the operator Fourier transform and not the quantum Fourier transform~\cite{babbush2017low}.
The $XY$ Hamiltonian on a ring is then exactly diagonalized as\cite{lieb_two_1961}
\bqa
H_{XY} \eq \sum_{k=1}^{\kappa} E_k f_k^\dag f_k 
\eqa
where the eigen-energies $E_k=2\cos(2k\pi/\kappa)$.
Replacing the number operator $f_{k}^{\dag}f_{k}$ with qubit operators $(1-\sigma_k^z)/2$, 
the Hamiltonian can be expressed as
\begin{align}
H_{XY}^{(k)} = \sum_{k=1}^{\kappa} E_k
\left( 1 - \sigma_{k}^{z}\right)/2\;\;
\end{align}
where the upper index $(k)$ is added as a reminder that we are in the momentum representation.
In this representation evolving $e^{-i\beta H^{(k)}_{XY}}$ involves only single-qubit $Z$-rotations.

The FFFT has emerged as a route to efficient simulation for fermions in tensor networks~\cite{Ferris14} 
and quantum circuits representing fermionic systems~\cite{Verstraete09, babbush2017low}.  
The circuit is constructed in a similar structure to the decimation-in time radix-2 classical Fourier transform and inherits the divide-and-conquer complexity.  
The FFFT circuit can be implemented with $\mathcal{O}(\mathrm{log}(\kappa))$~\cite{Ferris14} depth for a system with parallel arbitrary two-qubit interactions.  
For more realistic systems where only nearest-neighbor interactions are allowed fermionic swaps are required to swap the two modes together to perform the butterfly operation.  
This adds an additional overhead resulting in a $\mathcal{O}(\kappa \mathrm{log}(\kappa))$ circuit depth and $\mathcal{O}(\kappa^{2} \mathrm{log}(\kappa))$ total gate count~\cite{babbush2017low}.   The gate depth required to implement the FFFT was further improved to $O(\kappa)$ in ~\cite{Kivlichan_18} by using a Givens rotation network and requires only linear connectivity.

We also point out that the Givens rotation network is a powerful tool for state preparation for general quadratic Hamiltonians.  For pairing models, the linear depth network was used to prepare ground states~\cite{PhysRevApplied.9.044036}.  This initial state can be used in the context where the hard constraint is of the form that qubits must appear paired up.  We point this out as an example of how different flavor constraints can correspond to evolving a wide variety of constraint-preserving Hamiltonians.
\subsection{Linear depth simultaneous complete-graph mixer}\label{sec:circuit_complete}
We consider the simultaneous mixer for a node, $e^{-i\beta H_{{\tt complete}},v}$,
with $H_{{\tt complete},v}=\sum_{c<c'=0}^{\kappa-1} H_{XY,v,c,c'}$, 
which corresponds to a complete graph of variables corresponding to all colors for each vertex $v$, $\{x_{v,c}\}$.
Beyond a one-dimensional layout, the analytical solution to the $XY$ model is not known, 
therefore, exactly realizing the evolution of $XY$ model on a complete graph poses a challenge.
In this section we show that within the subspace of total $\ZtotV = \pm (\kappa -2)$ as in our case, 
when $\kappa=2^m$, this unitary can be exactly implemented in a circuit of depth $\kappa-1$, 
up to a constant factor accounting for breaking a generic two-qubit operator to 
any fixed universal set of single- and two-qubit operators.

We illustrate the process using $\kappa=4$ and then show the general formula.
For $\kappa=4$, we consider three partitions of the full set of colors: $\{\{0,1\},\{2,3\}\}$, $\{\{0,2\},\{1,3\}\}$ and $\{\{0,3\},\{1,2\}\}$, in the feasible subspace, we have
\bqa
& & \exp[-i\beta \sum_{c,c'\in[0,3]}{(XY)}_{c,c'}] =  \nonumber\\
& & \exp[-i\beta\big({(XY)}_{0,1}+{(XY)}_{2,3}\big)]   \nonumber\\
& &  \exp[-i\beta\big({(XY)}_{0,2}+{(XY)}_{1,3}\big)]   \nonumber\\
& &  \exp[-i\beta\big({(XY)}_{0,3}+{(XY)}_{1,2}\big)]\;,
\eqa
where for notational simplicity we use $(XY)_{j,j'}$ to refer to the XY Hamiltonian $H^{(XY)}_{j,j'}$ defined below Equation~\eqref{eq:parity}.
Note that this equivalence is approximate in general but exact if we consider only the action on the feasible subspace.
The fact that these partitioned operators commute in the feasible subspace can be easily verified mathematically.

The following perspective on the partitioning scheme allows us to derive a generalization for any $\kappa$. 
Consider an integer variable $x\in\{0,1,2,3\}$, 
in the one-hot encoding, in the feasible space, 
the $XY$ operation on a pair of qubits swaps the integer values the states represent.
For example $(XY)_{1,3}$ swaps between the variable taking value $1$ and taking value $3$.
Now consider the $2$-bit binary encoding of $x$: $x = 2^1 x_1  + 2^0 x_0$ where $x_0$ and $x_1$ are bits.
The swap between $\{0,1\}$ and $\{2,3\}$ corresponds to flipping the zero-th bit $x_0$.
The swap between $\{0,2\}$ and $\{1,3\}$ corresponds to flipping the first bit $x_1$.
The swap between $\{0,3\}$ and $\{1,2\}$ corresponds to flipping both bits $x_0$ and $x_1$.
Such operations can happen in any order without affecting the final value of $x$,
hence the corresponding partitioned mixers commute.

For a general $\kappa=2^m$, the partition can be read out taking the inverse of this process: 
all pairs involved in each $l$-bit flipping form a partition, for $l=1,\ldots,m$.
There are $\binom{m}{l}$ many $l$-bit flips, 
hence the total partitions $\sum_{l=1}^{m}\binom{m}{l}=2^m-1 = \kappa-1$.
Within each partition the pair-wise $XY$ operators commute and can be executed simultaneously. 
The simultaneous complete-mixer unitary can be accordingly executed in depth $\sim \kappa-1$. 

For example, partitions for $\kappa=8$ can be prescribed using the following equations.  Here we use $\tilde 0, \tilde 1, \tilde 2$ to represent bits in the binary encoding.  The left-hand side for each equation is a $l$-bit flip operation, all seven operations commute.  The right hand side is derived from reading off the effect of the operation on the numerical colors.  The right-hand side gives the corresponding partition for the $XY$ operators in the one-hot encoding, with the detailed procedure displayed in Table.~\ref{tb:demo_binary}.
\begin{widetext}
\bqa
I_{\tilde 2}I_{\tilde 1}X_{\tilde 0}
\eq (XY)_{01}+(XY)_{23}+(XY)_{45}+(XY)_{67}  
\nonumber\\
I_{\tilde 2}X_{\tilde 1}I_{\tilde 0} 
\eq (XY)_{02}+(XY)_{13}+(XY)_{46}+(XY)_{57} 
\nonumber\\
X_{\tilde 2}I_{\tilde 1}I_{\tilde 0} 
\eq (XY)_{04}+(XY)_{15}+(XY)_{26}+(XY)_{37} 
\nonumber\\
I_{\tilde 2}X_{\tilde 1}X_{\tilde 0} 
\eq (XY)_{03}+(XY)_{12}+(XY)_{47}+(XY)_{56} 
\label{eq:demo}\\
X_{\tilde 2}I_{\tilde 1}X_{\tilde 0}
\eq (XY)_{05}+(XY)_{14}+(XY)_{27}+(XY)_{36} 
\nonumber\\
X_{\tilde 2}X_{\tilde 1}I_{\tilde 0} 
\eq (XY)_{06}+(XY)_{17}+(XY)_{24}+(XY)_{35} 
\nonumber\\
X_{\tilde 2}X_{\tilde 1}X_{\tilde 0}
\eq (XY)_{07}+(XY)_{16}+(XY)_{25}+(XY)_{34}
\eqa
\end{widetext}
\begin{table}[htb]
\begin{tabular}{|l|l|l|l|}\hline
 $decimal$ &  one-hot&  Apply IXX & decimal \\\hline\hline
  0&  000&   011& 3  \\\hline
  1&  001&   010& 2  \\\hline
  2&  010&   001& 1  \\\hline
  3&  011&   000& 0  \\\hline
  4&  100&   111& 7  \\\hline
  5&  101&   110& 6  \\\hline
  6&  110&   110& 5 \\\hline
  7&  111&   100& 4 \\\hline
\end{tabular}
\caption{Demonstration of deriving the partition
corresponding to operator $IXX$ on the binary encoding, Eq.~\eqref{eq:demo}}.
\label{tb:demo_binary}
\end{table}
\subsection{On realistic layout of physical qubits}\label{sec:compilation}
To achieve the above derived circuit depth for simultaneous ring and complete-graph mixers
requires physical qubits to have a particular connectivity.
For example, all-to-all connectivity for the simultaneous complete-graph mixer.
For a physical-qubit-layout of lower connectivity, SWAP operations may be necessary to enable the pairwise operations~\cite{Venturelli18_Compiling, guerreschi2018qaoa}.

We first note that in the feasible space, 
an $XY$ operation $\exp[-i\beta (XY)_{j,j'}]$ and a SWAP$_{j,j'}$ executed consecutively are equivalent 
to a $XY$ of different parameter: 
\begin{align}
\label{eq:xy_swap}
e^{-i\beta (XY)_{j,j'}} \text{SWAP}_{j,j'}
= i e^{-i (\beta+\frac{\pi}{2}) (XY)_{j,j'}}\;.
\end{align}
or, in matrix form
\begin{align}
\begin{pmatrix}
1 & 0 & 0 & 0 \\
0 & -i \sin{\beta} & \cos{\beta} & 0 \\
0 & \cos{\beta} & -i \sin{\beta} & 0 \\
0 & 0 & 0 & 1 \\
\end{pmatrix}
\end{align}
This relation can be explored in circuit compilation to achieve a circuit on 
physical qubits of lower depth.  

For example in the $\kappa=4$ example for simultaneous complete-graph mixer in Sec.~\ref{sec:circuit_complete}
if the physical qubits form a ring $1-2-3-4-1$,
the three partitions can be executed in the following order: 
\begin{align*}
\{\{0,1\},\{2,3\}\}\\
\{\{0,3\},\overline{\{1,2\}\}}\\
\{\{0,2\},\{1,3\}\}
\end{align*}
where the pair with an overline indicates a SWAP is combined in the $XY$ mixing in the fashion of Eq.~\eqref{eq:xy_swap}.
This SWAP enables the operation of the last partition, and the whole circuit is of depth $3$, the same as in the case of a complete graph.

For a general $\kappa$ simultaneous mixer, this feature helps scheduling but cannot avoid SWAPs completely because the partitions need to be executed in a specific order.

On the other hand, executing the pairwise $XY$ unitary in any given order would give a valid partitioned mixer, though not equivalent to the simultaneous one.  Exploiting the feature Eq.~\eqref{eq:xy_swap} combined with a bubble sort scheme, we can completely avoid SWAPs and implement a partitioned complete-graph mixer in linear depth.

The same SWAP network circuit architecture, derived from a fermionic simulation perspective, can be used to implement the phase-separator Hamiltonian with no Trotter error~\cite{Kivlichan_18, crooks2018performance}.  Viewing the qubits as an array of $\kappa$-sites the parallel bubble sort algorithm implements 
a swap network in $\kappa$-depth 
where every element (qubit) of the array is swapped passed each other once. 

Since all terms in the phase-separator commute, there is no incurred Trotter error. The swap interaction can be efficiently combined with the evolution of a $e^{-i\theta ZZ}$ 
nearest-neighbor interaction by adding an $e^{-i\theta Z}$ in between the second 
and third CNOT (denoted CX below) in the SWAP decomposition~\cite{Venturelli18_Compiling}.  

Explicitly,
\begin{align}
\mathrm{SWAP}_{0, 1}  e^{-i\theta ZZ_{0, 1}} =& \mathrm{CX}_{0, 1} \mathrm{CX}_{1, 0} \mathrm{CX}_{0, 1} e^{-i\theta ZZ_{0, 1}} \\
=&\mathrm{CX}_{0, 1} \mathrm{CX}_{1, 0}.e^{-i\theta Z_{1}} \mathrm{CX}_{0, 1}
\end{align}
can be used as the swap interaction and simultaneously evolve a local $ZZ$-interaction term.  
For any encoding model that increases the dimensions of the graph, such as in the case where one-hot-encoding is used, 
simulating the interaction term removes the necessity of encoding techniques such as minor embedding 
or classical logical encoding~\cite{Lechnere15, Rocchettoe1601246}.

\section{Simulation Results}\label{sec:simulation}
In this section, we present the results of numerical simulations of QAOA applied to the max-$\kappa$-Colorable-Subgraph problem.  We first compare the performance of the $XY$ mixer to that of the $X$ mixer with penalty.  We then more deeply explore the behavior of $XY$ mixers, looking at general features of their performance on small hard-to-color graphs, and comparing complete-graph $XY$ mixers against ring $XY$ mixers.

To acquire a good set of QAOA parameters, stochastic optimizer is needed, in Appendix.~\ref{sec:landscape} we show rugged landscape with local optima in the parameter space that would cause problem for deterministic optimizing methods like gradient descend.  We instead use basin-hopping with Broyden-Fletcher-Goldfarb-Shanno (BFGS) algorithm to obtain (sub)optimal parameters.

\subsection{The death of \texorpdfstring{$X$}{X}-mixer}
\label{sec:triangle}
We use a simple example, 2-coloring and 3-coloring of a triangle to demonstrate
the performance comparison of $XY$ and $X$ mixers.

\begin{figure}[htbp]
\subfloat[]{\hspace{-2mm}
\includegraphics[width=.25\textwidth]{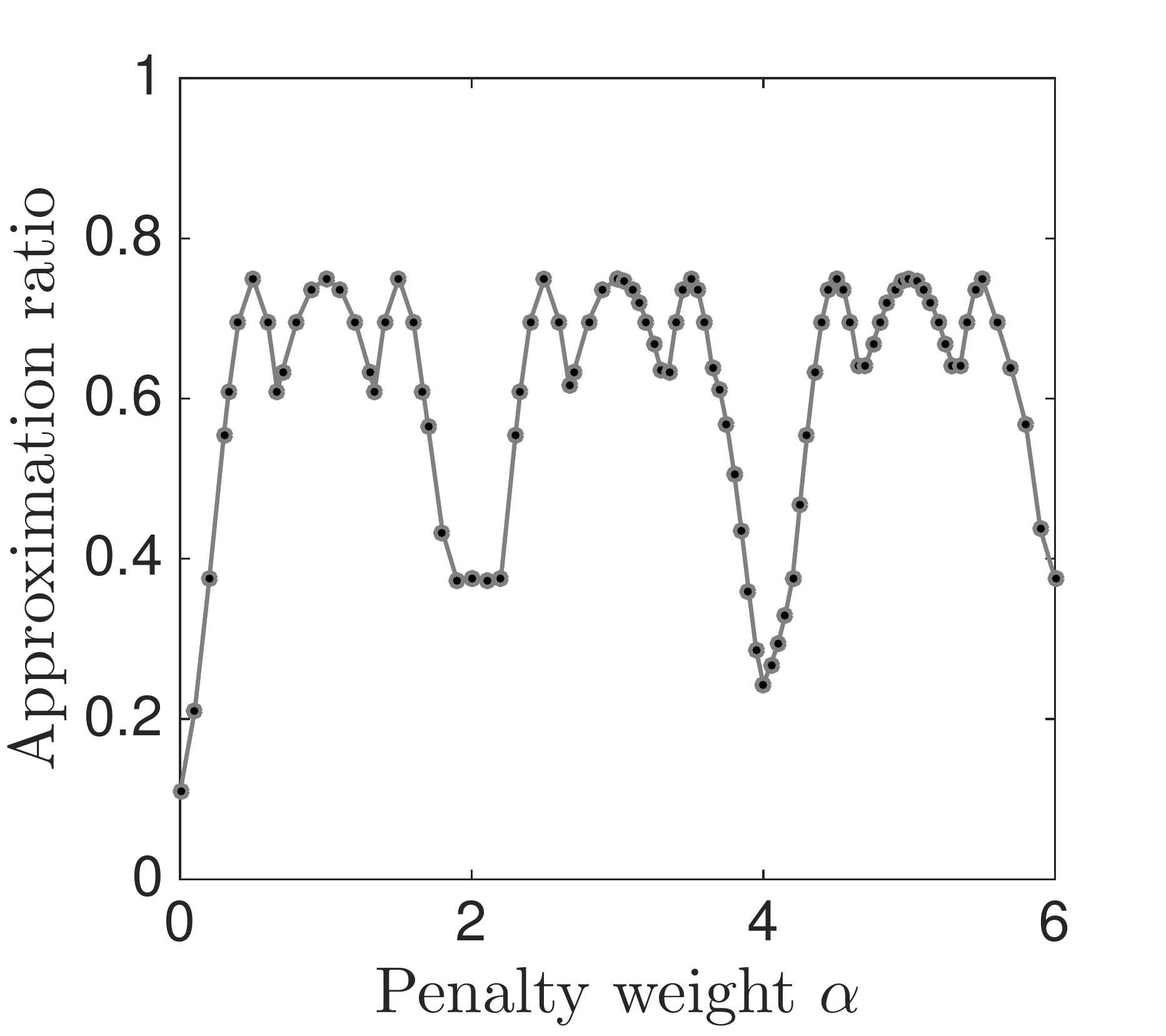}
\label{fig:triangle_k2p1}
}
\subfloat[]{\hspace{-4mm}
\includegraphics[width=.25\textwidth]{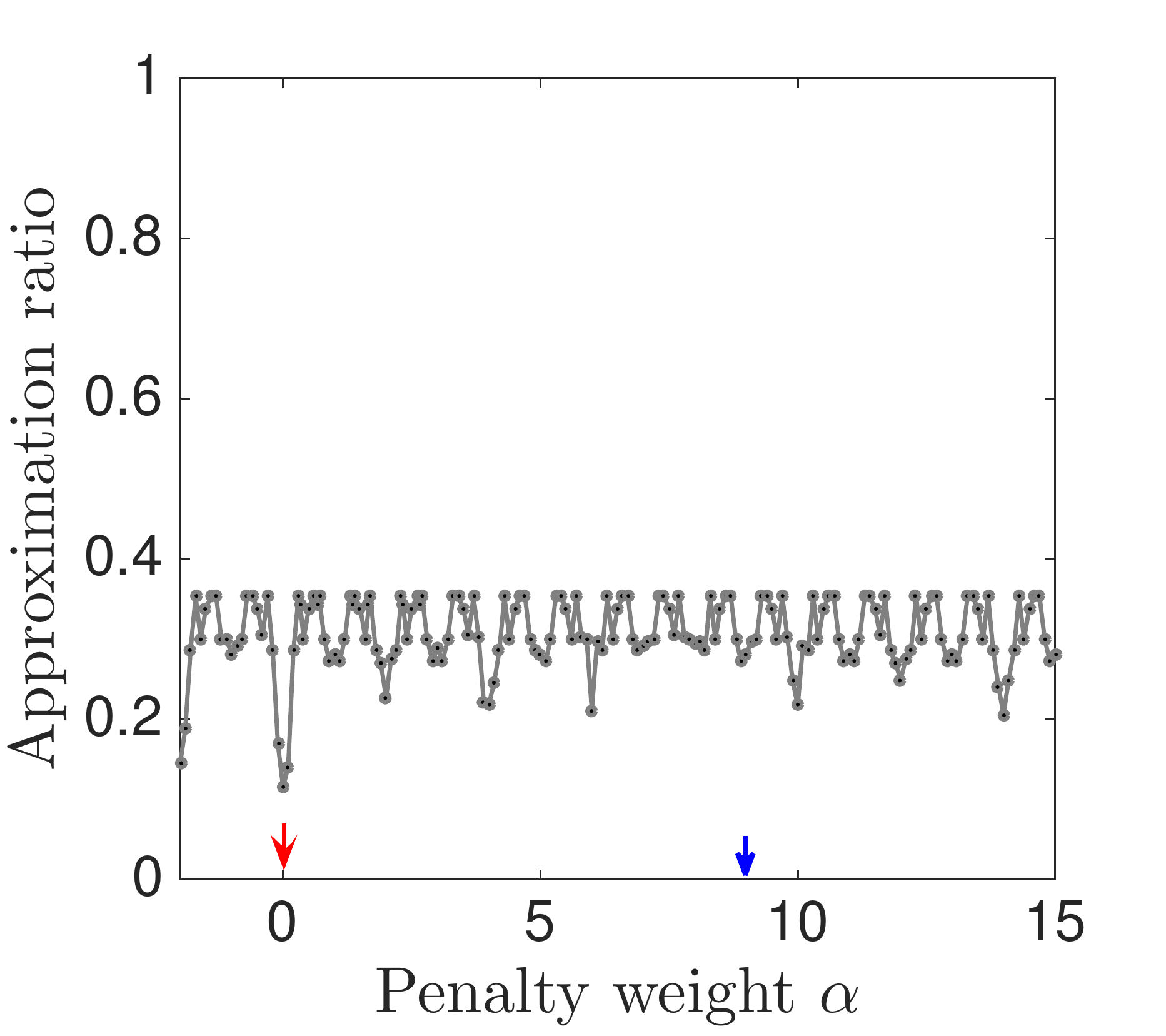}
\label{fig:triangle_k3p1}
}
\caption{
(a) 2-coloring (b) 3-coloring of triangle with level 1 $\QAOA$.
The highest approximation ratio across the parameter sets ($\gamma_1, \beta_1$)
is plotted versus the penalty weight $\alpha$. 
The red arrow at $\alpha=0$ indicates the minimum penalty that guarantees the optimal state being the optimal state in the feasible subspace,
and the blue arrow at $\alpha=9$ indicates the penalty value that guarantees separation between energies feasible and infeasible states.
} 
\label{fig:triangle_penalty}
\end{figure}

Note that the penalty weight $\alpha$ in general affects the performance of the algorithm. 
In Fig.~\ref{fig:triangle_penalty} we show that for 2-coloring 
the approximation ratio optimized over the parameter set $(\gamma,\beta)$ for each penalty weight $\alpha$.
The best approximation ratio, $r$, takes value $0.75$ while with $XY$ driver QAOA$_{p=1}$ gets $r=1$.

In Fig.~\ref{fig:triangle_penalty}, while the penalty strength has an effect on the behavior of level 1 QAOA, there appears to be no clear intuition for choosing a good value.  
In particular, 
the minimum penalty that guarantees the optimal state being the optimal state in the feasible subspace, indicated by the red arrows on the plots, does not stand out,
nor does the penalty value that guarantees separation between energies feasible and infeasible states, indicated by the the blue arrows.
This supports our argument in Sec.~{sec:penalty} that 
the role of energy gap plays no clear role in QAOA.

\begin{figure}[htbp]
\subfloat[]{\hspace{0mm}
\includegraphics[width=.24\textwidth]{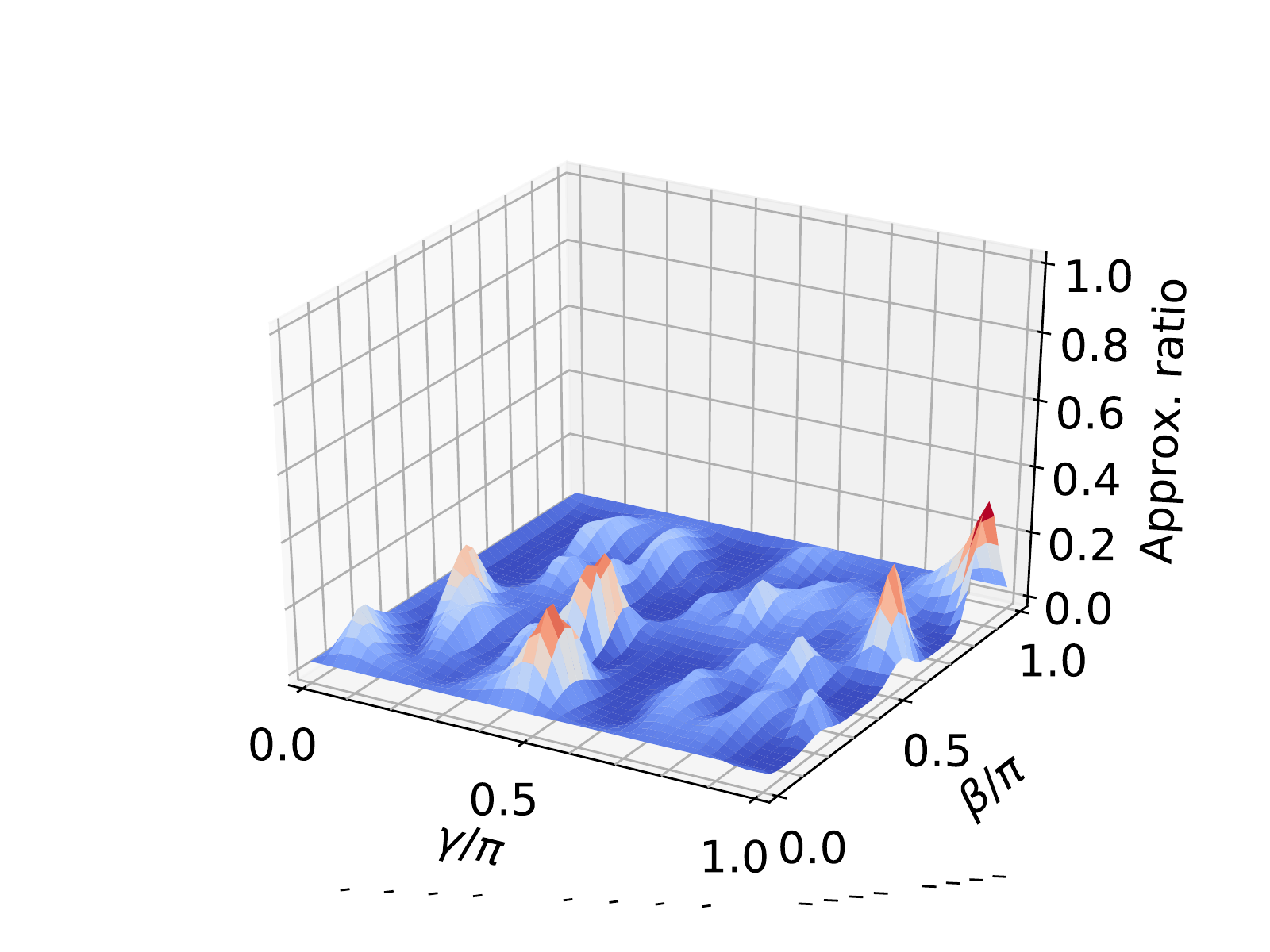}
\label{fig:triangle_X}
}
\subfloat[]{\hspace{-4.7mm}
\includegraphics[width=.24\textwidth]{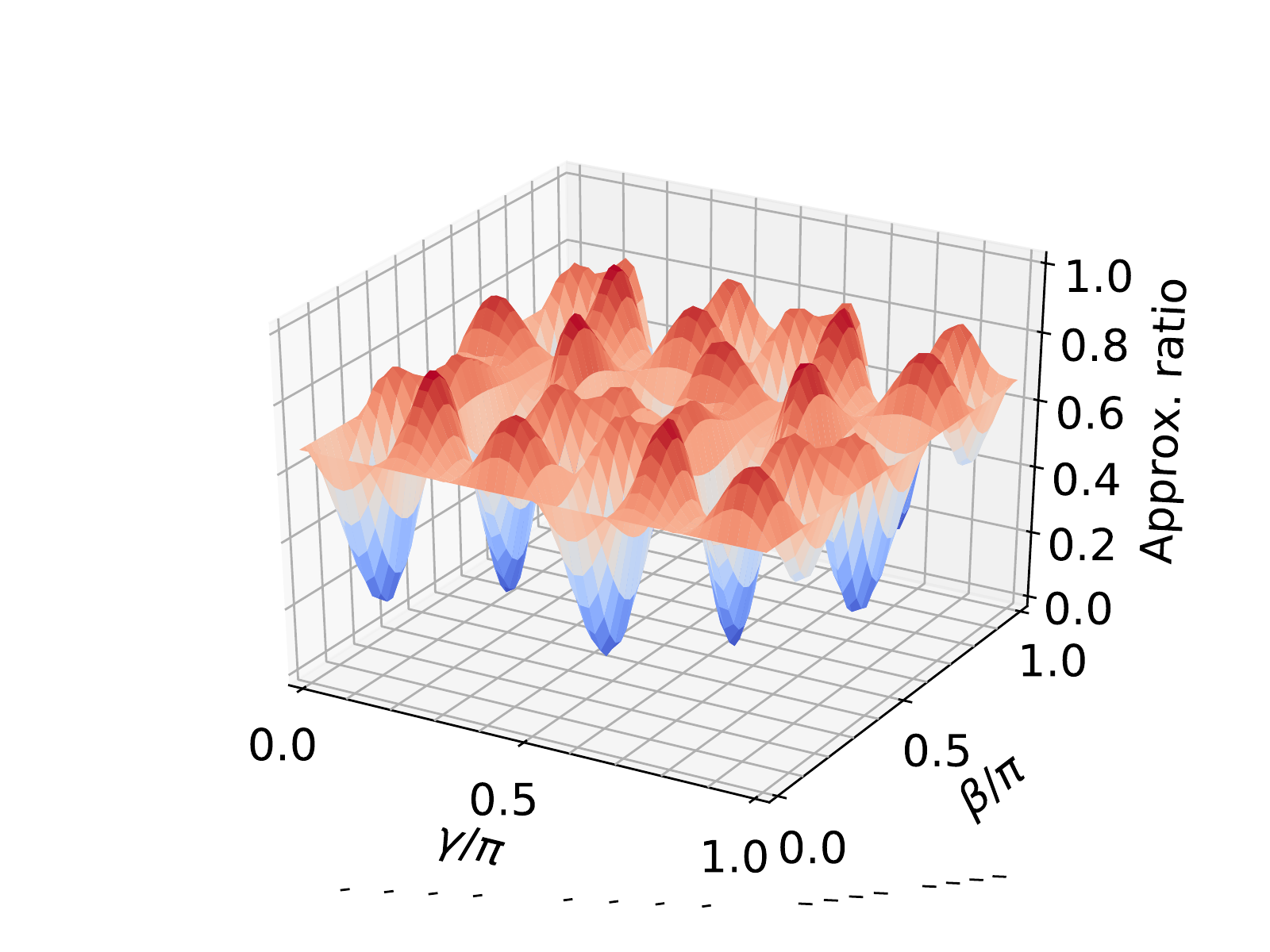}
\label{fig:triangle_XY}
}
\caption{
Numerical results for level 1 $QAOA$ on the problem of 3-coloring of a triangle graph.
(a) using $X$ mixer 
along with phase-separating Hamiltonian, Eq.~\eqref{eq:phaseHam} where the penalty weight is taken to be the numerically determined optimal value $\alpha^*=1.7$.}
(b) using the $XY$ mixer with W-state being the initial state. 
\label{fig:triangle_p1}
\end{figure}

For 3-coloring, in Fig.~\ref{fig:triangle_p1} 
we plot how the approximation ratio varies 
in the 2-dimensional $(\gamma,\beta)$ space, for using the $X$ mixer and for using the $XY$ mixer.
While with the $X$ mixer the QAOA$_{p=1}$ gives approximation ratio $\sim 0.2$ across the parameter value range,
with the $XY$ mixer parameter values that correspond to $\sim 0.8$ can easily be found.
This example thus shows significant performance advantage in using the $XY$ as compared with the $X$ mixer.

\subsection{Small and hard-to-color graphs}
\label{sec:small_and_hard}
\begin{figure}[htbp]
\begin{center}
\includegraphics[width=0.8\columnwidth]{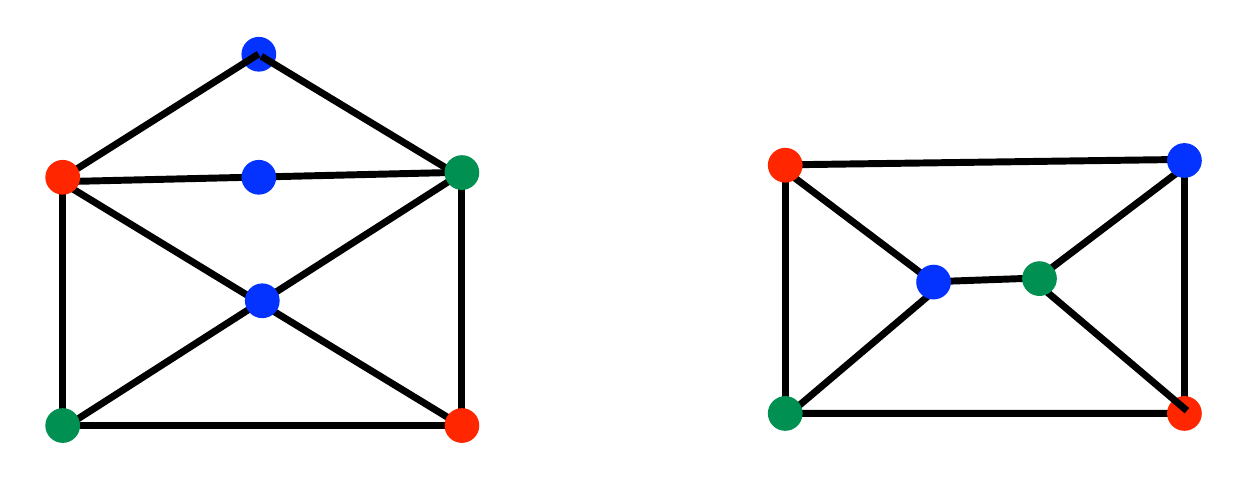}
\end{center} 
\caption{
\label{fig:envelop_prism}
The two small and hard-to-color graphs: Envelope and Prism. A valid 3-coloring is shown on each graph.
}
\end{figure}

For a fixed classical algorithm, 
a \emph{slightly-hard-to-color} graph is a graph for which the algorithm will sometimes yield the optimal solution.
Similarly, a \emph{hard-to-color} graph is one such that the chosen algorithm never yields the optimal solution. Two examples are the Envelope and the Prism graphs,\cite{Kosowski2008} sketched in Figure.~\ref{fig:envelop_prism}.  The Prism graph is the smallest slightly-hard-to-color graph for the smallest-last(SL) sequential coloring method and 
the Envelope graph is the smallest hard-to-color graph for the largest-first(LF) sequential method.
Note that these classical algorithms are aiming to compute the chromatic number, 
while in this paper we focus on finding the maximal colorable subgraph.
Although finding the max-colorable subgraph could serve as a subroutine for determining chromatic numbers,
the chromatic number can also be directly attacked by QAOA using a much more complex mixer.\cite{hadfield2017quantum} 
Nevertheless we are not aiming at doing side-by-side comparison of quantum and classical algorithms,
and will use these small graphs only as a proof-of-principle demonstration of the QAOA with $XY$ mixers.

\subsubsection{Performance of QAOA with the simultaneous ring mixer}
\begin{figure}[htbp]
\begin{center}
\includegraphics[width=\columnwidth]{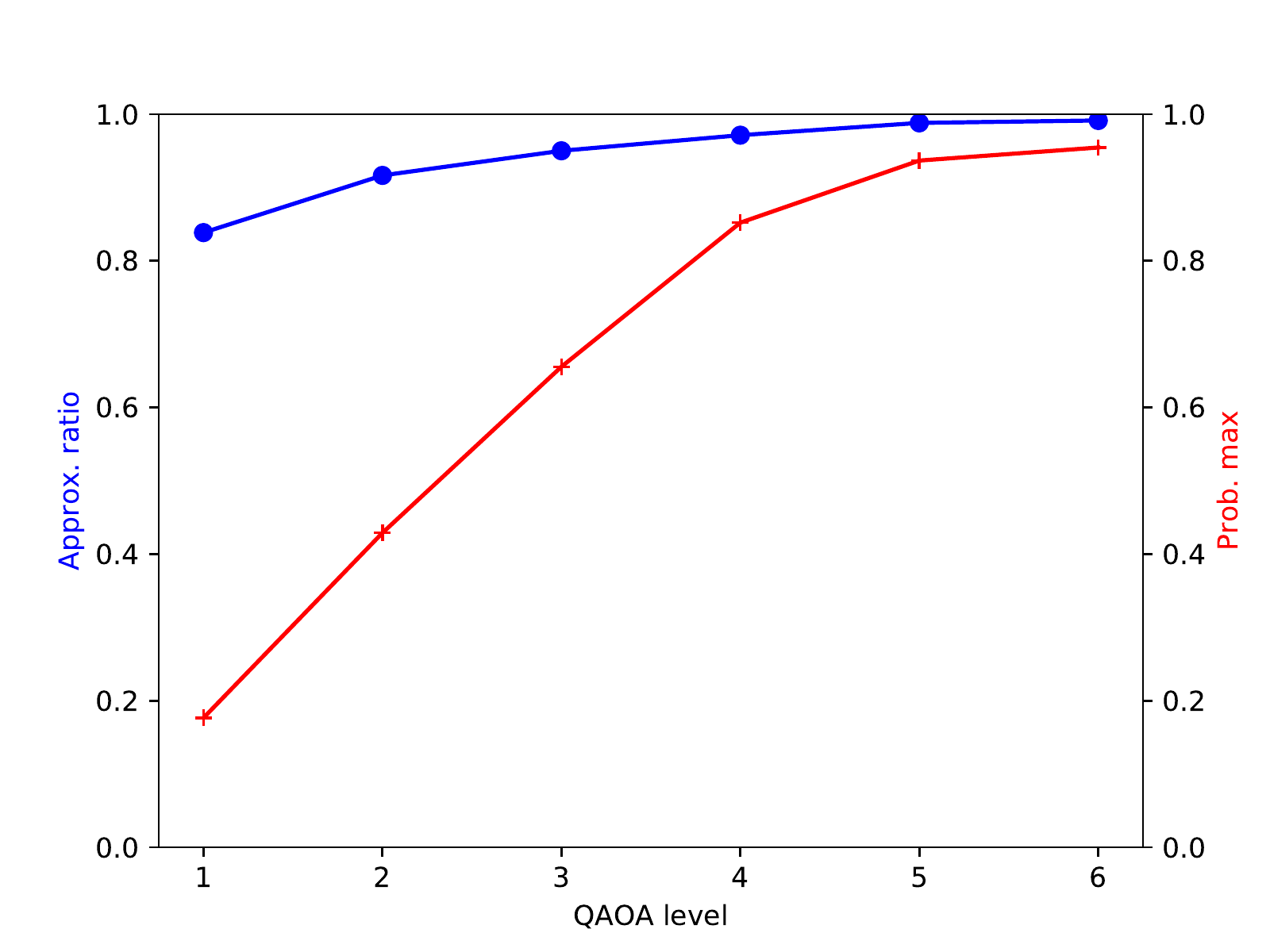}
\end{center} 
\caption{
\label{fig:prism_level}
The Prism graph.
Dots are approximation ratios and crosses are the expected probability of getting the optimal coloring.
For each QAOA level, 
results are shown at the (sub)optimal angles
resulted from a basin-hopping search.
}
\end{figure}

With the simultaneous ring mixer,
Figure.~\ref{fig:prism_level} shows the results for QAOA levels 1 to 6.
For each level, the W-state is used as initial state,
and stochastic search \zw(basin-hopping with BFGS) is performed to optimize
the expected value of the cost Hamiltonian over the angle sets. 
The approximation ratio corresponding to the optimal expectation value is plotted as filled circles.
Even at level one, the approximation ratio takes a high value 0.8, 
and this value quickly approaches 1 as the level increases.
Furthermore, for each level, we computed the probability of getting the actual optimal solution (a valid 3-coloring) upon measurement.  
At level one, this probability is slightly lower than 0.2, and quickly goes above 0.6 at level-3,
which implies that repeating the experiment 3 times, one will find a valid coloring with probability $>0.9$.

\subsubsection{Effect of initial states}\label{sec:initial}
The W-state -- as both an even superposition of all feasible classical states, 
and the ground state of the simultaneous ring mixer -- is a natural candidate for the initial state for QAOA.
It involves multiple two-qubit gates to prepare. 
An easier-to-prepare state for each vertex can be defined via a randomly-assigned coloring 
(feasible but not necessarily optimal), $\ket {\psi_C}$, i.e., a randomly drawn bit string of Hamming weight one.
Preparing such a state involves only $n$ single-qubit gates.

\begin{figure}[htbp]
\begin{center}
\includegraphics[width=\columnwidth]{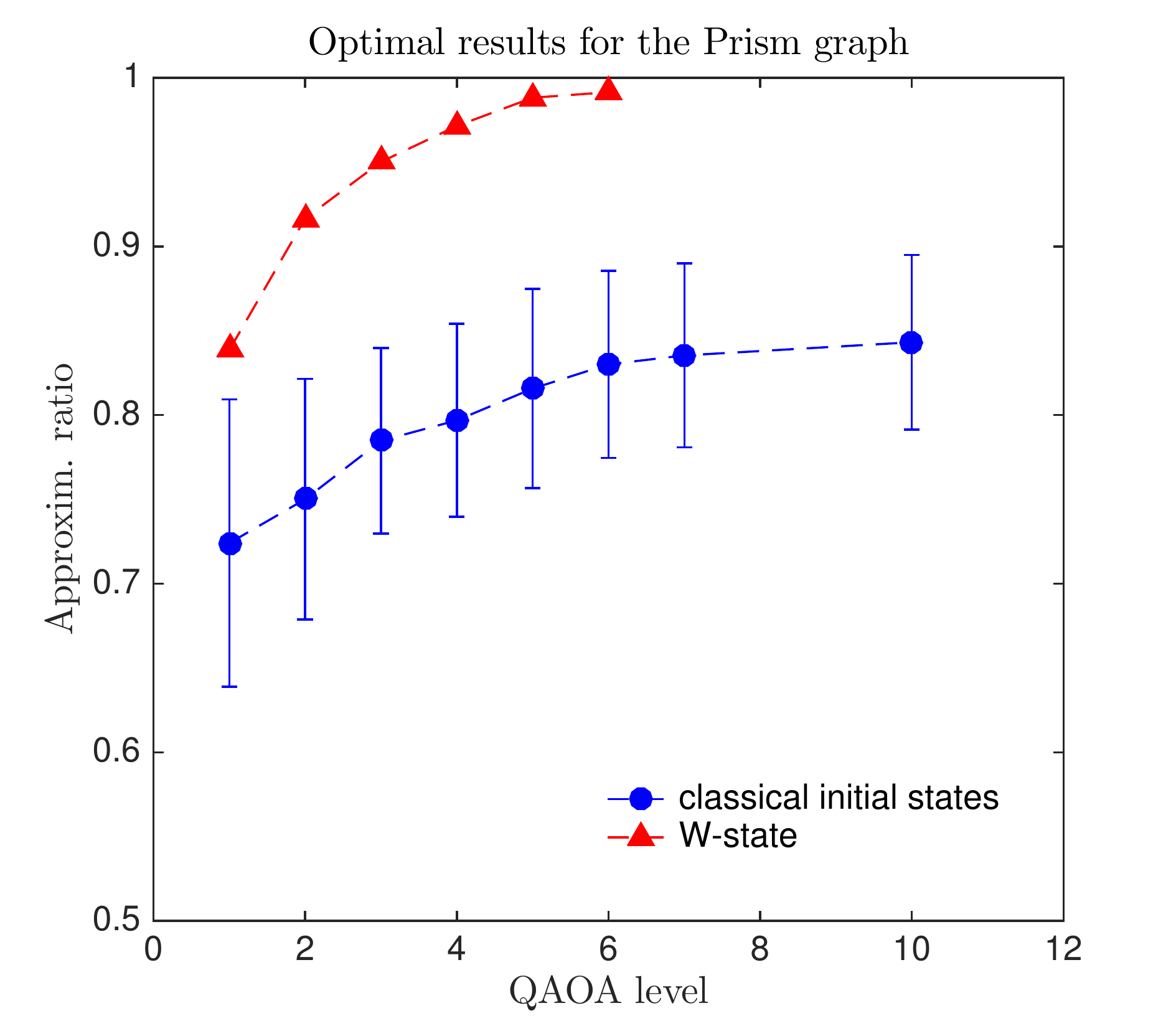}
\end{center} 
\caption{
\label{fig:prism_ini_cmp}
The Prism graph, the expected value of QAOA optimized over the angle sets.
Triangles show the results with W-state as initial states.
Circles show the results with a feasible classical initial state, averaged over the set of all feasible classical states, 
the error bar is the standard deviation.
For each initial state, optimization over angles are derived from a basin-hopping search.
}
\end{figure}

We study both initial states for the prism graph with simultaneous ring mixer.
For level-1 QAOA, the best achievable optimization ratio (optimized over all angle sets $(\beta,\gamma)$) 
for W-state is higher than the classical Hamming weight 1 state $\ket \psi_C$.
Notice that for $\ket \psi_{C}$, the phase-separating unitary commutes with the density matrix of the state, hence has no effect to the state evolution.  
As a result, the whole circuit for level-1 QAOA is equivalent to applying the mixing unitary followed by measurement. 
We further simulated higher levels, 
and in Figure.~\ref{fig:prism_ini_cmp} show the performance of QAOA with the W-state versus a classical state as initial state.
We found that with the classical initial state, the performance of QAOA is 
significantly lower than using the W-state as initial state.
Even at level 10, $r_{classical}$ is still lower than $r_{W}$ for level-1.
Moreover, the approximation ratio with classical initial state shows a tendency toward saturation around level 10 
-- this could either be the nature of the algorithm, or due to increasing difficulty in finding the global optimum in the parameter subspace as the level increases, which poses another practical consideration for application.
(Note that due to the optimization over parameter space for each initial state, 
the average over classical initial state is not equivalent to prepare the initial state in a mixed state for the ensemble).

Because our simulation is noise-free,
due to ergodicity, in the limit of $p\to\infty$ the optimal performance 
should be independent of the initial state.
But for practical implementation on a near-term hardware 
where noises accumulates fast with circuit depth, 
such medium-level QAOA behavior is of high relevance.
In Appendix~\ref{sec:W} we survey methods to generate quantum circuits for preparing W-states.
It is shown that with certain methods it can be generated with $O(\kappa)$ CNOT gates.
The overall performance of QAOA will be a tradeoff between the extra effort in preparing W-state 
and the damage that comes with circuit depth.

\subsection{Benchmarking graph sets}\label{sec:benchmark}
To better understand the behavior of these QAOA graph-coloring algorithms, we make use of the sets of all $\kappa$-chromatic graphs of size $n$ as the benchmarking sets 
for the $XY$ mixers under consideration.
See Table~\ref{tb:benchmark_sets} for the number of instances in each benchmarking set.

\begin{table}[htb]
\begin{tabular}{|l|l|l|}\hline
 $\chi$ &  $n$&  No. graphs \\\hline\hline
  3&  5&    12  \\\hline
  3&  6&    64  \\\hline
  3&  7&    475  \\\hline
  4&  6&    26  \\\hline
  4&  7&    282  \\\hline
  5&  7&    46  \\\hline
  6&  7&    5  \\\hline
\end{tabular}
\quad
\begin{tabular}{|l|l|l|}\hline
 $n$&  $\kappa$&  No. graphs \\\hline\hline
   4&  4&  6  \\\hline
   4&  6&  6  \\\hline
   4&  8&  6  \\\hline
\end{tabular}
\caption{Left: Benchmarking graph sets:
each row indicates all $\chi$-chromatic graphs of size $n$,
and we solve the problem of $\kappa$-coloring of such graphs choosing $\kappa=\chi$.
Right: Benchmarking graph sets II for examining the 
simultaneous vs partitioned ring mixers on different ring sizes:
Each row indicates all connected graphs of size $n$,
and we solve the problem of $\kappa$-coloring of such graphs.
Because the total number of qubits is $n\kappa$, which is the limiting factor to the simulation, 
we limit to small $n$ to see $\kappa$ varying up to $8$.
}
\label{tb:benchmark_sets}
\end{table}

\begin{figure}[htbp]
\begin{center}
\includegraphics[width=\columnwidth]{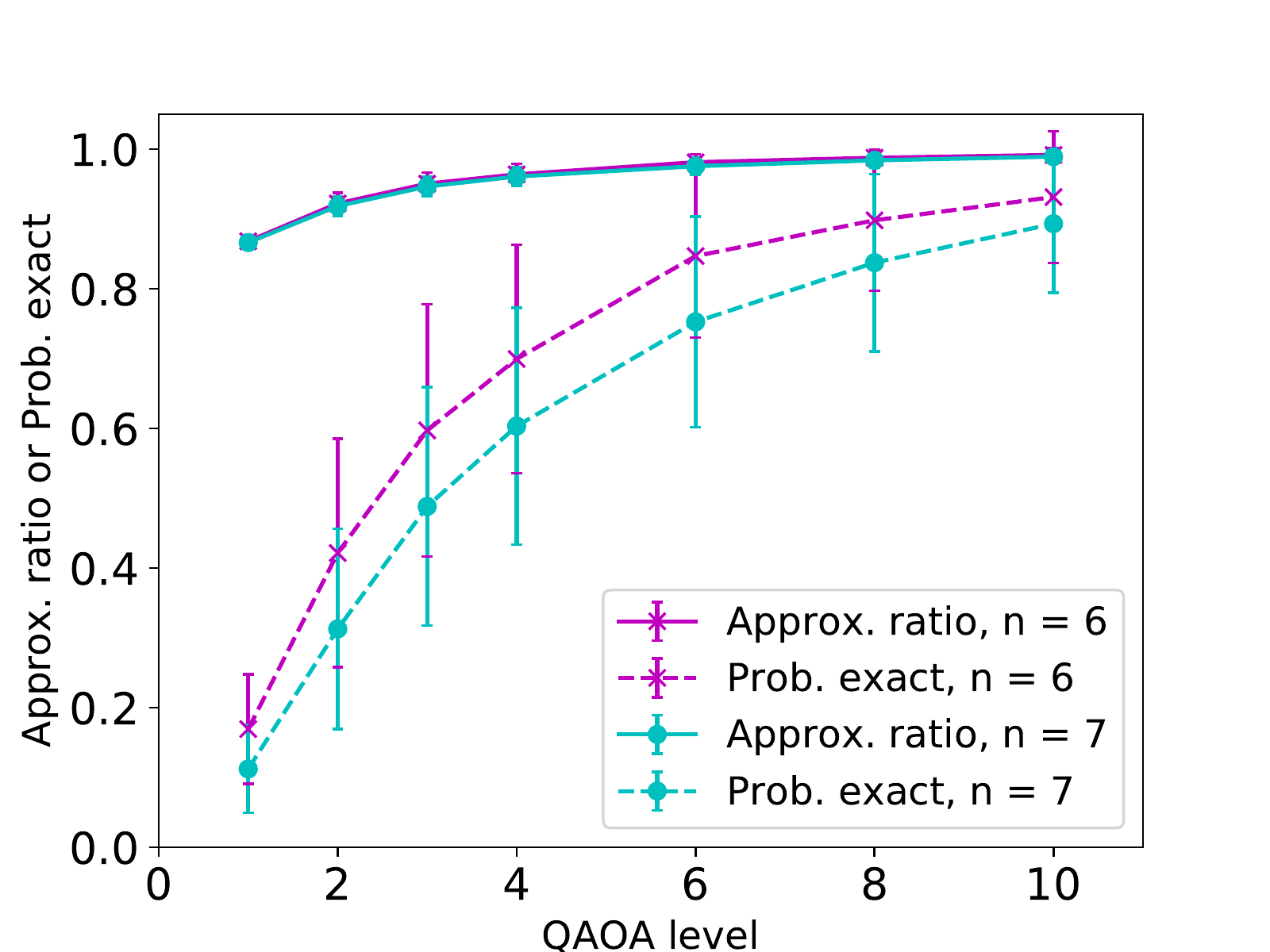}
\end{center} 
\caption{
\label{fig:cmp_size}
Approximation ratio (solid lines) and probability to exact solution (broken lines) for QAOA with ring simultaneous mixer. $n=6$ (crosses) vs $n=7$ (filled circles).
}
\end{figure}

\begin{figure} [htbp]
\begin{center}
\subfloat[ QAOA level-2 ]
{
\includegraphics[width=0.5\columnwidth]{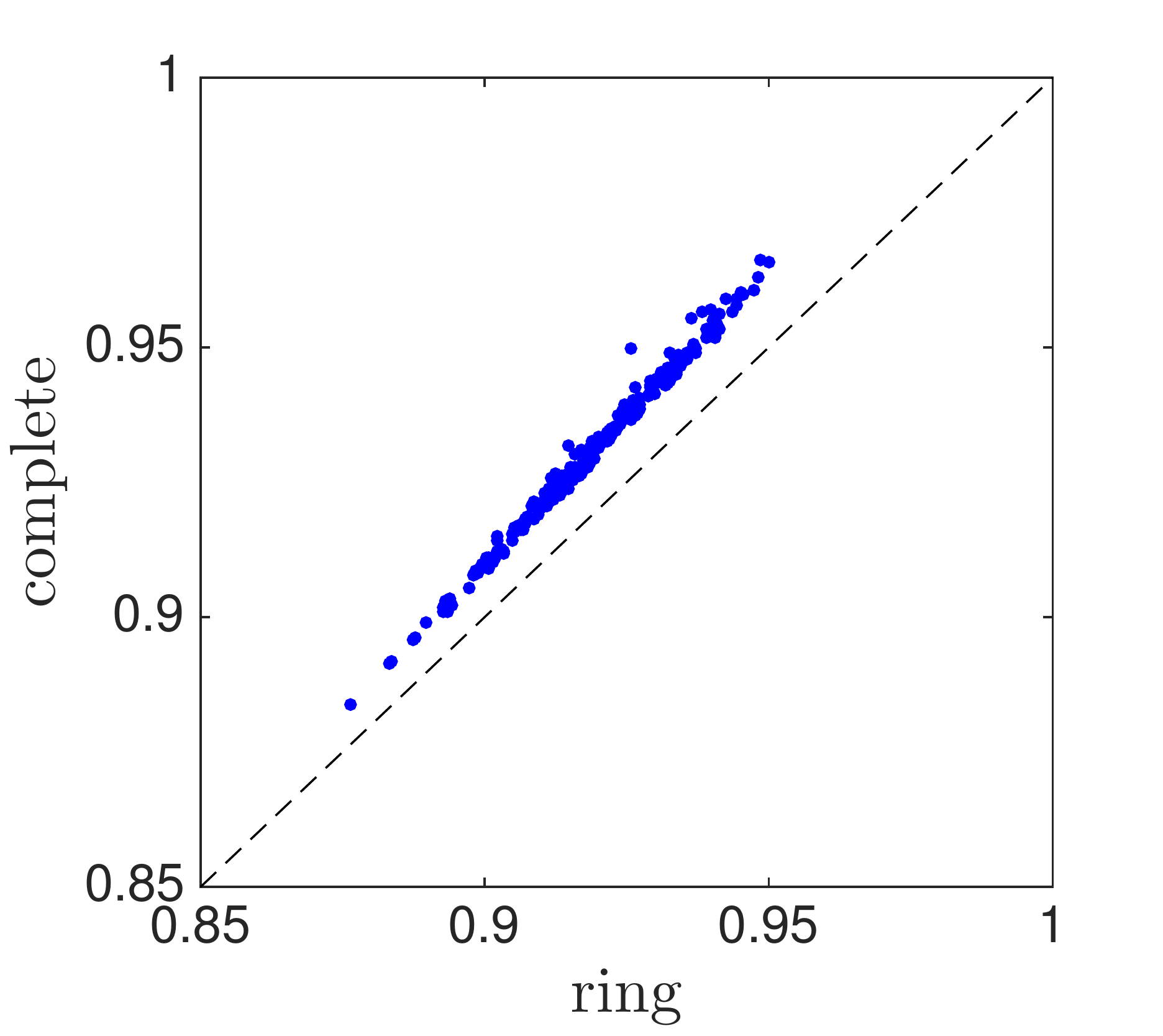}
\label{fig:chi_4_n_7_p_2}
}
\subfloat[ QAOA level-8 ]
{
\includegraphics[width=0.5\columnwidth]{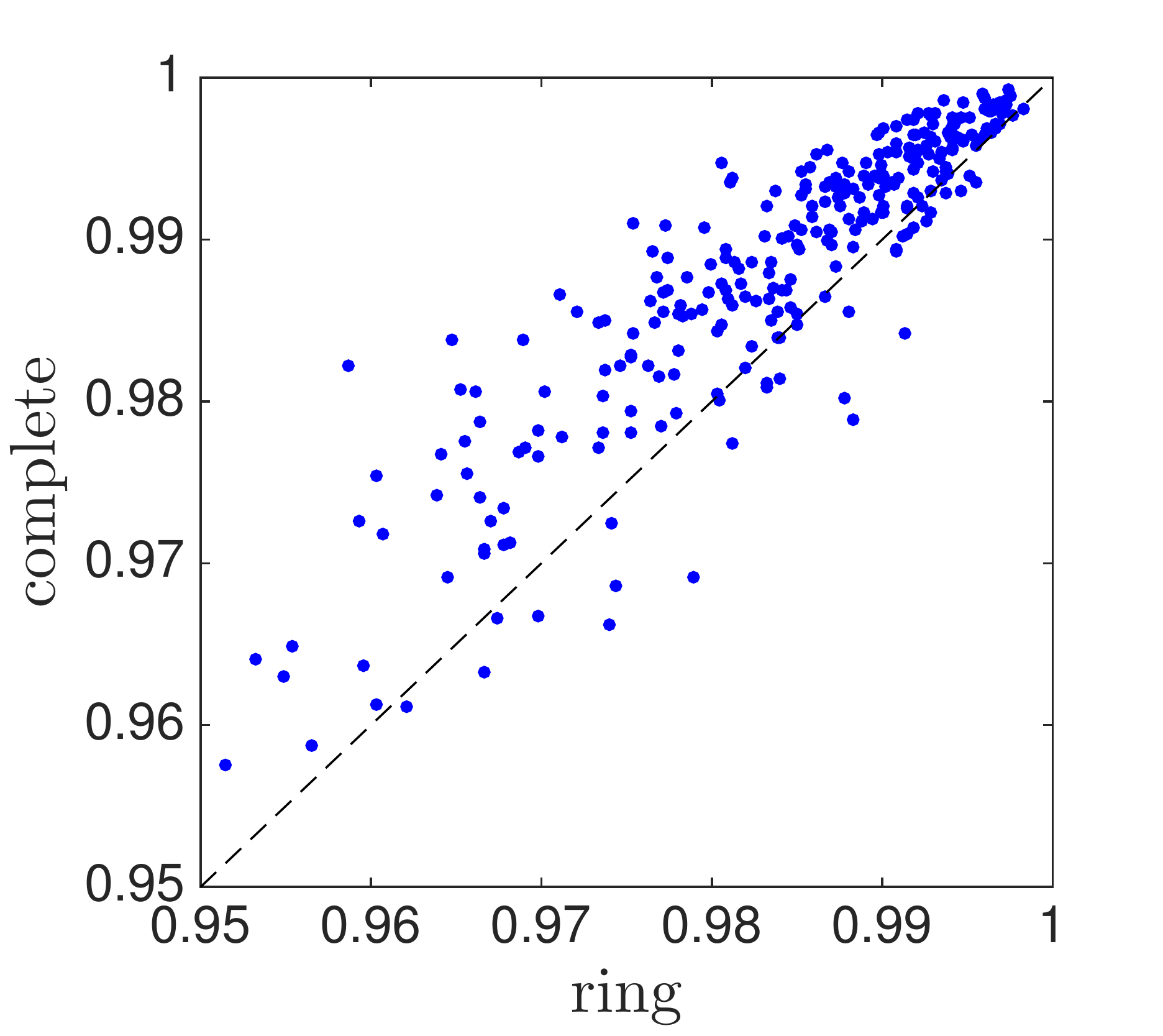}
\label{fig:chi_4_n_7_p_8}
}
\end{center} 
\caption{
\label{fig:ring_vs_complete}
QAOA with simultaneous mixers.  Performance comparison between ring and complete-graph mixers applied to the same graph coloring problems.
The axes show approximation ratio achieved using the labeled mixer type.
Scatter plot shows the results for
$4$-coloring of all connected chromatic-$4$ graphs
of size $n=7$.
In (b), for better visibility, an outlier data point at (ring = 0.95, complete = 0.9) is not shown in the plot.
}
\end{figure} 


\subsubsection{Approximation ratio and probability-to-optimal-solution}
\label{sec:benchmark_mean}
Using $W$-state as the initial state, for simultaneous ring and complete-graph mixers, 
the mean and median of the approximation ratio as well as 
the probability-to-optimal-solution
is evaluated across problem sets.

The following observations have been made on the typical performance for each problem set.

\paragraph{Consistent performance over instances.}
For all problem sets, the approximation ratio and the probability-of-optimal-solution curves as a function of the QAOA level 
are highly consistent across graphs, bearing the same shape for the Prism and Envelope graphs.
For each problem set, the approximation ratio showed very little deviation from the mean (demonstrated by the small error bars in Figure.~\ref{fig:cmp_size}).

\paragraph{Larger graphs are harder to color.}
As expected, for the same $\kappa$, as $n$ increases, 
the performance of QAOA with the same type of mixer decreases, 
see Figure.~\ref{fig:cmp_size} for comparison of the simultaneous ring-mixer for $n=6$ and $n=7$.

\paragraph{Complete-graph mixer is better than the ring mixer.}
For the same problem size $n$, the simultaneous complete-graph mixer 
demonstrates better performance than the simultaneous ring-mixer in QAOA levels from 1 to 10.  See the scatter plot for QAOA level-2 and level-8 in  Figure~\ref{fig:ring_vs_complete}.  For small QAOA levels, this advantage is uniform cross instances for smaller levels, as shown in~Figure~\ref{fig:ring_vs_complete} (a) for level-2 where for all 282 instances the complete mixer generates higher approximation ratio.  The advantage is decreasing as QAOA level increases, see comparison of (a) and (b).  This is possibly due to the approximation ratio getting close to 1.  We also speculate that the QAOA level where this closeup happens would vary with, $\kappa$, the number of colors. 

\paragraph{Similar performance between the simultaneous and parity mixers for small $\kappa$.}
We also study $\kappa$-coloring of all connected graphs (regardless of chromatic number) of size $n=3,~4$,
with varying $\kappa$ to compare simultaneous vs partitioned ring mixers on different ring sizes.
Since for $\kappa=4$, the simultaneous and the parity mixers are equivalent,
we will need to go for higher $\kappa$ to examine the difference,
however, numerical power is limited by the number of qubits $n\kappa$,
we thus examined two classically trivial cases, $n=4,~\kappa=6$ and $8$ (trivial coloring exists).
Both approximation ratio and probability of exact solution are high due to the small problem size,
and no noticeable difference is observed between the performance of partitioned and simultaneous mixers.
Extensive studies on larger problem sizes are needed to further evaluate these two types of mixers.

\subsubsection{Typical solution upon measurements}
\begin{figure}[htbp]
\begin{center}
\includegraphics[width=.9\columnwidth]{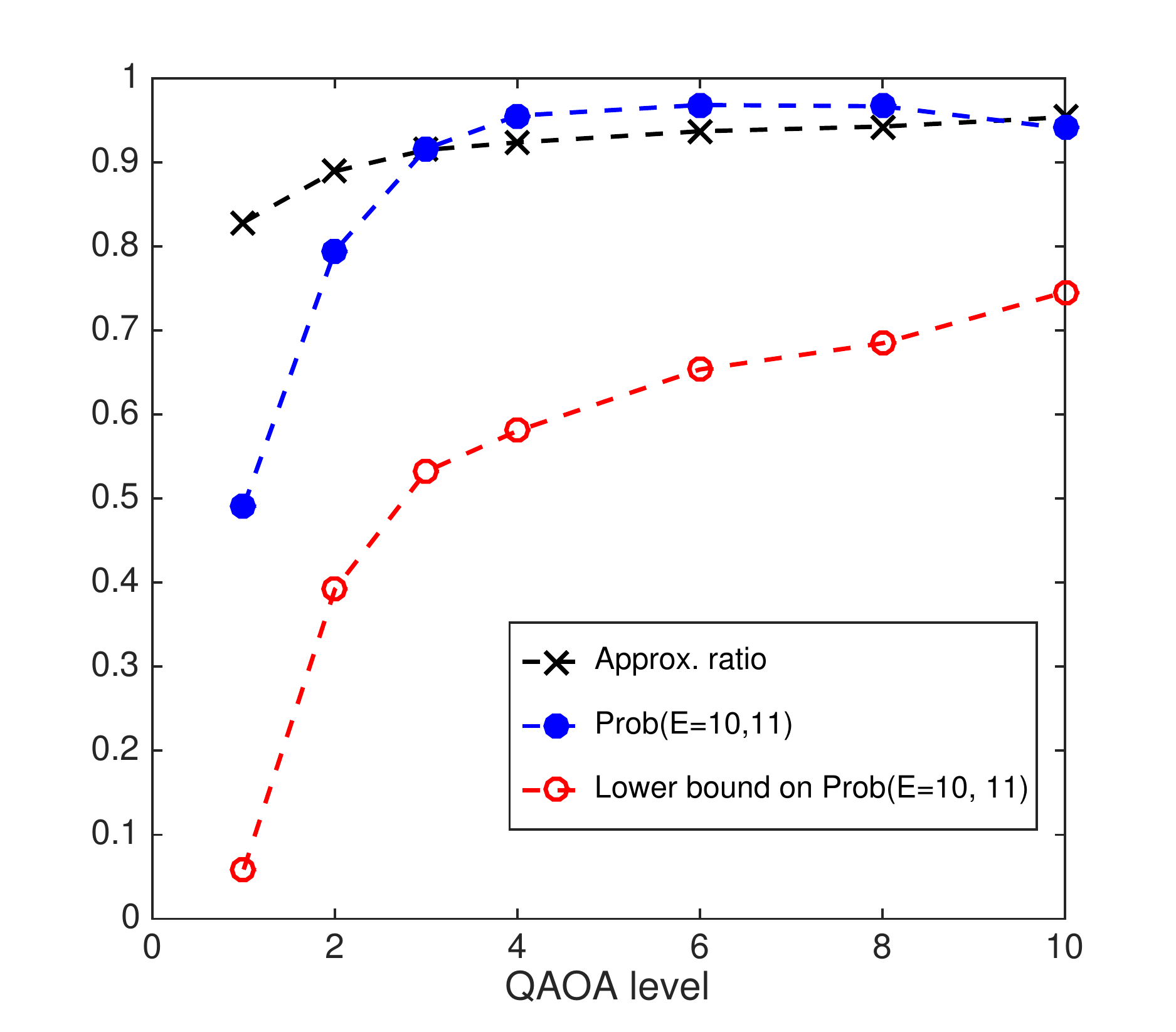}
\end{center} 
\caption{
\label{fig:envelope_bound_prob}
$3$-coloring of Envelope graph (11 edges).  QAOA with simultaneous ring-mixer. 
For each QAOA level, the probability of getting the top two highest approximate results (cost 11 and 10)
is shown in comparison to the bound given by Eq.~\eqref{eq:bound} with the observed approximation ratio as parameter.
}
\end{figure}

Note that our optimization over the set of angles is designed to maximize the expected value,
and the high approximation ratio discussed in Section~\ref{sec:benchmark_mean} is also about the expected value.
For approximate optimization, the expectation value of the approximation ratio is not the whole story.
One also cares about the probability of getting the optimal or near-optimal states upon measurement.
We apply the argument and analysis on the tail probability in Sec.~\ref{sec:QAOA_framework}, 
Eq.~\eqref{eq:bound}, on the case of $3$-coloring of the Envelope graph (11 edges), 
and show in Figure~\ref{fig:envelope_bound_prob} the theoretical lower bound in probability of getting a solution with costs $10$ or $11$, i.e., 
the valid coloring or only one improperly-colored edges.
The true probability from evaluating the wavefunction is shown for comparison.
For QAOA level three and up, the bound inferred from the approximation ratio 
gives us confidence that with greater than $50\%$ probability we will get the optimal or the next best solutions.

Viewing the QAOA as an exact solver, 
as observed in the case of small hard-to-color graphs, 
for the benchmarking problem sets, we also see that as $p$ increases, 
along with the increase in $r^*$, there is a more drastic increase in the probability of getting a optimal solution.
In Fig.~\ref{fig:cmp_size} we also plot the mean  prob-to-optimal-solution as $p$ changes,
with error bars indicating the standard deviation over the graphs in the set.
\section{Conclusion}\label{sec:conclusions}
Exploring the range of applications of the QAOA provides insight into what can be achieved with near-term quantum resources.  While the general search for applications of QAOA is important, the detailed specification of the algorithm can be the difference between success and failure when running the algorithm on a real device.  These details can change if the gate-model computing substrate is switched--e.g. switching from superconducting qubits to ions.  For example, constant factors in circuit depth gained by switching gate sets can drastically change circuit depth and thus the success probability of the overall algorithm.  

In this work we explored applying QAOA to optimization over $\kappa$-ary variable sets.  Our representative example of this optimization was the max-$\kappa$-colorable-subgraph problem on small hard-to-color graph problems.  We numerically demonstrate and theoretically motivate that the $XY$-mixer Hamiltonian is a natural choice for this particular set of problems.  Part of the numerical analysis was providing circuit implementations for the phase-separator and the $XY$-mixer under various qubit topologies.  Though there is higher implementation cost of the $XY$-mixer in comparison to the standard $X$-mixer, the benefits of eliminating a penalty term and restricting dynamics to the feasible space can potentially outweigh the linear-depth implementation cost. 

Along with the circuit analysis the bound on tail effects based on the mean value we provide in Eq.~\eqref{eq:bound}
suggests that a high mean value is sufficient to guarantee performance without having to analyze the variance of the distributions produced by QAOA. 

This work establishes the possibility of using more sophisticated drivers in a QAOA framework for naturally including constraints.
We expect this analysis is helpful for near-term experimental validations of the QAOA algorithm and, hopefully, inspires alternative
constraint encodings that would lower the representational cost of real-world optimization problems.

\section{Acknowledgements}

ZW thank enlightening discussions with all Quail members, Zhang Jiang, and Sergey Knysh.  We are grateful for support from NASA Ames Research Center, and from the NASA Advanced Exploration systems (AES) program and the NASA Transformative Aeronautic Concepts Program (TACP), and also for support from the AFRL Information Directorate under grant F4HBKC4162G00. ZW is also supported by NASA
Academic Mission Services (NAMS), contract number
NNA16BD14C.

\appendix
\section{ Proof on finite tail probabilities}\label{app:bound}
In what follows, we will take $\{a_{j}\}_{j=0}^{K}\subset \mathbb{R}$ to be a strictly ordered finite set, i.e. $a_{0} < a_{1} < \dots < a_{K}$.  Suppose that $p_{j}$ is the probability of a random variable $X$ having outcome $a_{j}$, with $\sum_{j=0}^{K}p_{j} = 1$.  We will assume below that we know the values $\{a_{j}\}$ and the mean $\mu$ of this probability distribution, but not the entire distribution itself.
\begin{lemma}
	Given $\mu = \sum_{j=0}^{K}p_{j}a_{j}$, with $\mu \geq a_{l}$ for some $0\leq l\leq K$, 
	\begin{align}
		\Pr(X > a_{l}) \geq \frac{\mu-a_{l}}{a_{K}-a_{l}} = 1-\frac{a_{K}-\mu}{a_{K}-a_{l}}.\label{eqn:fixedMeanTailProb}
	\end{align}
\end{lemma}
\begin{proof}
	We will prove the bound using the method of Lagrange multipliers.  To ensure the probabilities are nonnegative, we will represent them as squares: $p_{j} = q_{j}^{2}$.  Then the Lagrangian is
	\begin{align}
		\mathcal{L} = \sum_{j=l+1}^{K}q_{j}^{2} + \lambda\left(\sum_{j=0}^{K}q_{j}^{2} - 1\right) + \gamma\left(\sum_{j=0}^{K}q_{j}^{2}a_{j} - \mu\right).
	\end{align}
	Differentiating and setting the derivatives equal to zero, we find the conditions
	\begin{align}
		2q_{j}\left(\delta_{>l}(j) + \lambda + a_{j}\gamma\right) = 0,
	\end{align}
	i.e.,
	\begin{subequations}
	\begin{align}
		\text{for } j\leq l:& & q_{j} & = 0  \qquad\text{or} & \lambda + a_{j}\gamma = 0\\
		\text{for } j> l:& & q_{j} & = 0  \qquad\text{or} & 1 + \lambda + a_{j}\gamma = 0.
	\end{align}
	\end{subequations}
	
	First, consider the case that $q_{j}=0$ for all $j > l$.  Then $p_{j} = q_{j}^{2} = 0$ for all $j > l$, so the only way $\mu \geq a_{l}$ can be satisfied is if $\mu = a_{l}$.  Then $\Pr(X > a_{l}) = \sum_{j=l+1}^{K}p_{l} = 0 = \frac{\mu-a_{l}}{a_{K}-a_{l}}$, so \eqref{eqn:fixedMeanTailProb} is satisfied.  Similarly, if $q_{j} = 0$ for all $j\leq l$, then $\Pr(X > a_{l}) = 1$, so this case represents the maximum, rather than the minimum of $\Pr(X > a_{l})$, and in any case, \eqref{eqn:fixedMeanTailProb} is satisfied.
	 
	Now, if $q_{m} \neq 0$ and $q_{n}\neq 0$ for $0\leq m< n \leq l$, then it follows that $\lambda = \gamma = 0$ and therefore $q_{j} = 0$ for all $j > l$, so that, as just argued, \eqref{eqn:fixedMeanTailProb} is satisfied.  If, on the other hand, $q_{m} \neq 0$ and $q_{n}\neq 0$ for $l<m<n\leq K$, then it follows that $\gamma = 0$ and $\lambda = -1$, so that $q_{j} = 0$ for all $0\leq j\leq l$, and therefore , \eqref{eqn:fixedMeanTailProb} is satisfied.
	
	What remains is the case that exactly one $q_{m}\neq 0$ for $0\leq m\leq l$ and exactly one $q_{n}\neq 0$ for $l<n\leq K$ and all other $q_{j}$ are zero.  Then $\Pr(X > a_{l}) = p_{n}$, $p_{m} = 1-p_{n}$ and $\mu = p_{m}a_{m} + p_{n}a_{n} = (1-p_{n})a_{m} +p_{n}a_{n}$.  Solving for $p_{n}$, we find
	\begin{align}
		\Pr(X > a_{l}) = p_{n} = \frac{\mu-a_{m}}{a_{n}-a_{m}} = 1-\frac{a_{n}-\mu}{a_{n}-a_{m}}.
	\end{align}
	It is easily seen that this expression for $p_{n}$ decreases as $a_{n}$ increases and decreases as $a_{m}$ increases, so the minimum $p_{n}$ obtained in this way is when $m = l$ and $n= K$:
	\begin{align}
		\Pr(X > a_{l}) = p_{n} = \frac{\mu-a_{l}}{a_{K}-a_{l}} = 1-\frac{a_{K}-\mu}{a_{K}-a_{l}},
	\end{align}
	from which it follows that \eqref{eqn:fixedMeanTailProb} is satisfied.
\end{proof}

\section{ Rugged landscape }
\label{sec:landscape}
\begin{figure}[htbp]
\begin{center}
\includegraphics[width=\columnwidth]{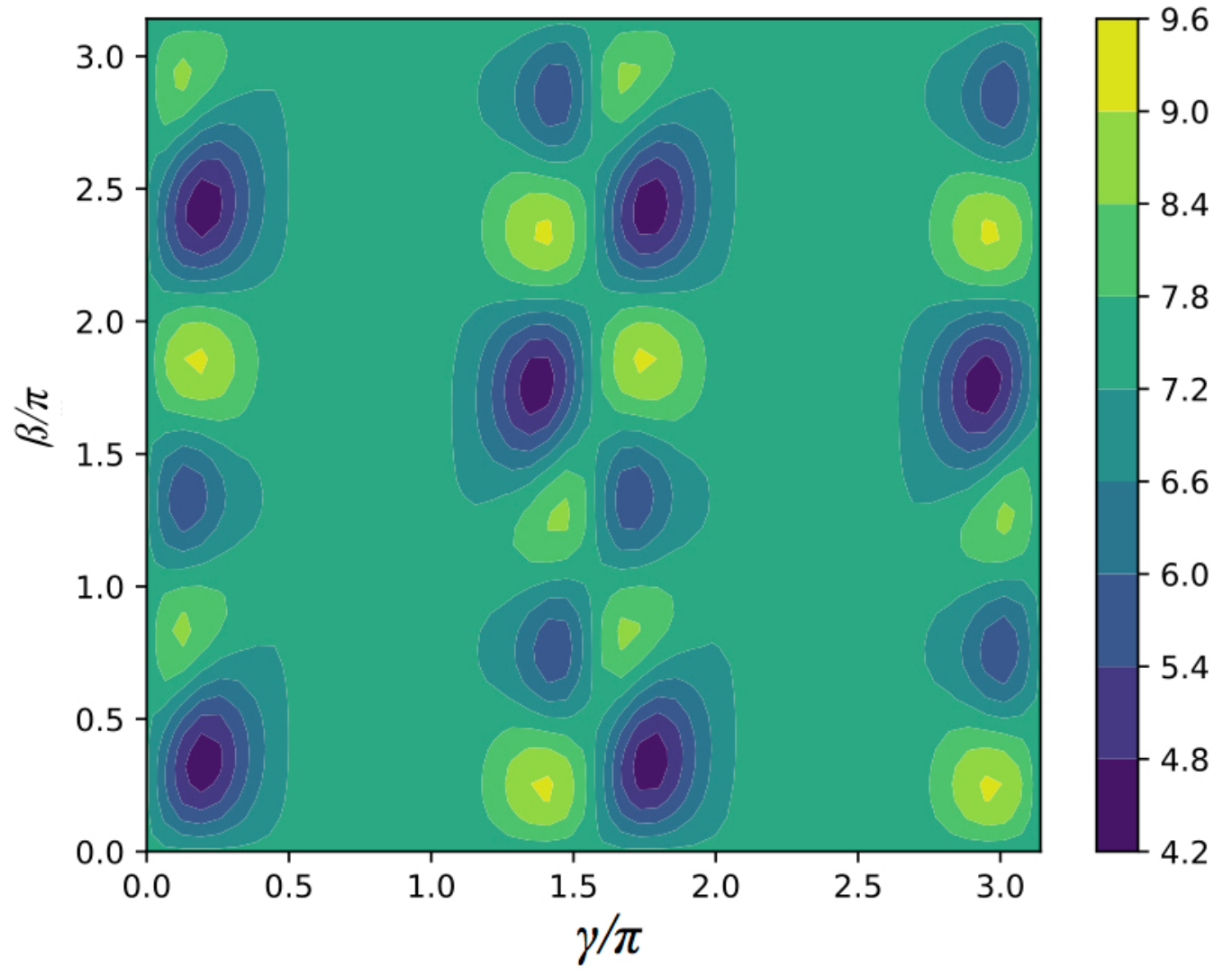}
\includegraphics[width=\columnwidth]{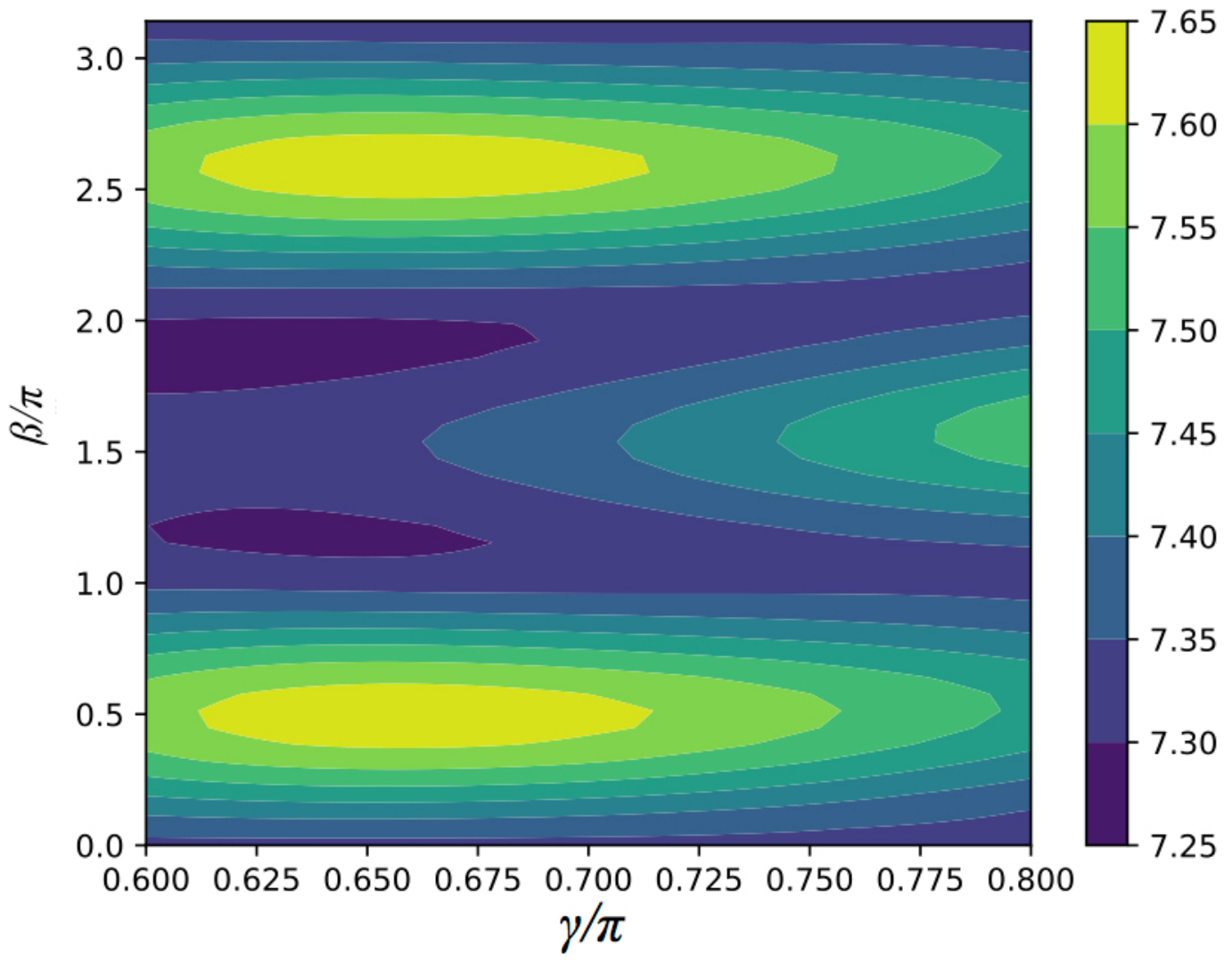}
\end{center} 
\caption{
\label{fig:landscape}
Landscape for level-1 QAOA for the envelope graph. Top: Full landscape.  Bottom: Zoom-in of the seemingly flat area that contains local maxima.
}
\end{figure}

The landscape of the parameter space plays an important role in quantum control.
In the case of MaxCut for a ring graph (equivalent to binary encoding for a 2-coloring of the ring), 
in Ref.~\cite{wang2017quantum} it has been observed that the landscape contains only global maximum.
In the current case of QAOA with $XY$ driver on graph coloring,
we notice that even for level-1 QAOA, the control landscape is rugged and contains local maximum.
In Fig.~\ref{fig:landscape} the landscape for the Envelope graph is plotted, and the bottom panel reveals local optima.
Stochastic optimization is therefore needed to perform parameter search.

\section{ W-state generation}\label{sec:W}
The $W$-state is a well-known multipartite-entangled state that plays an important role in quantum information theory.
Here, we survey methods to produce a generalized W-state using quantum gates.  
\subsection{Sequential generation of W-state}
\begin{figure}[htbp]
\begin{center}
\includegraphics[width=\columnwidth]{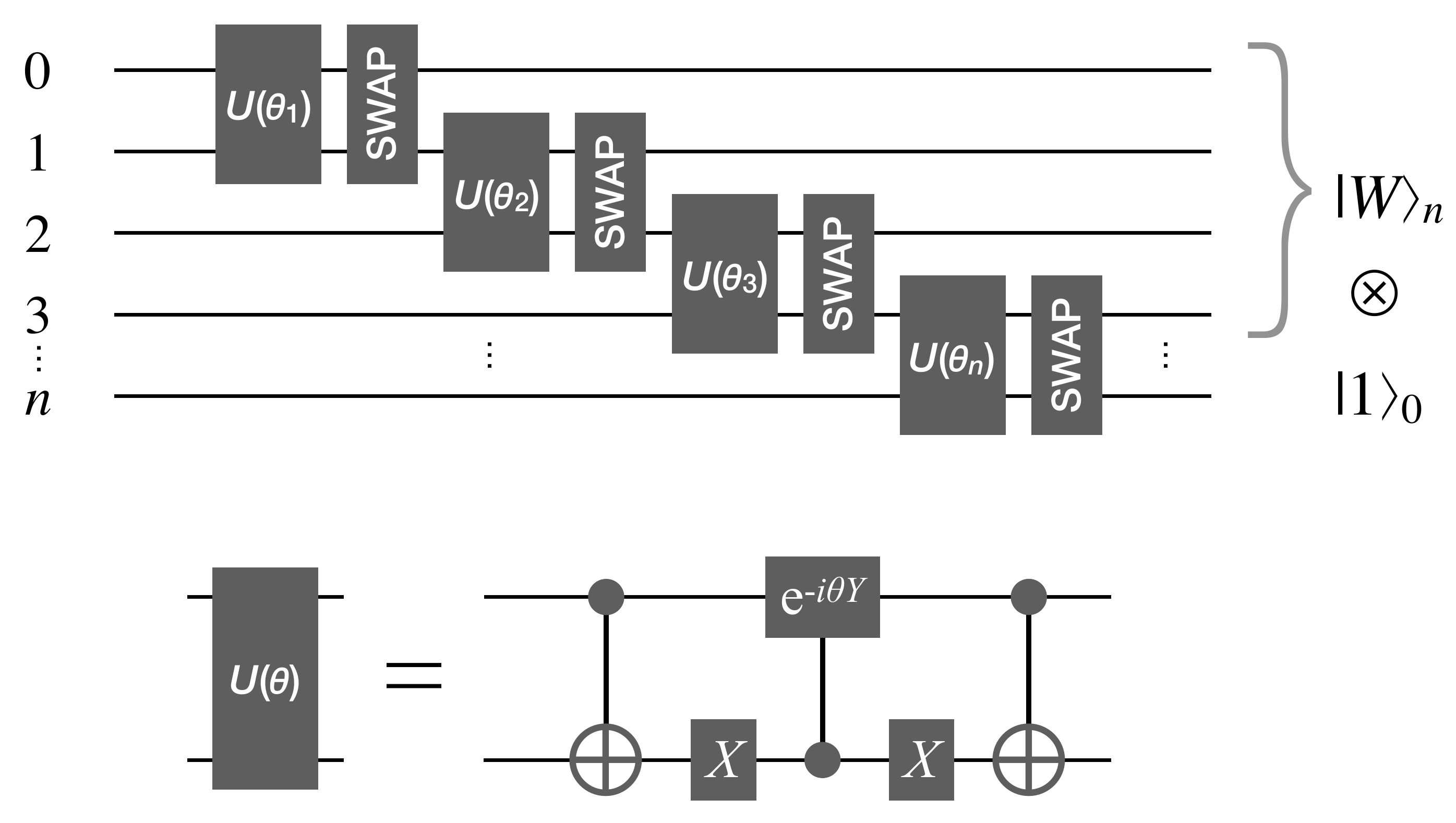}
\end{center} 
\caption{
\label{fig:w_circuit}
Circuit for generating the $\ket W$ state for arbitrary $n$.  
SWAP gates are added for illustration on 1D architecture of physical qubits. 
Qubit $n$ is discarded at the end.
}
\end{figure}

In Ref.~\cite{Schon07sequential},
it was shown a W-type state, which is any state living in the subspace spanned by states corresponding to Hamming-weight-1 bit strings,
can be generated using an auxiliary qubit by sequentially entangling it with each qubit.
Here we detail the case for the W-state using this method.
Consider an auxiliary qubit $q_0$ and an $n$-qubit register ($q_1$ to $q_n$),
initialized in the tensor product state
$\ket 0 \otimes {\ket 0}^{\otimes {n}}$.
Entanglement between $q_0$ and $q_j$ is introduced
by unitary 
\begin{align}
U_{0,j}(\theta_j,\phi_j) &=\ketbra{01}{01}+\ketbra{10}{10}\nonumber\\
& +c_j \ketbra{00}{00}+ c_j\ketbra{11}{11} \nonumber\\
& + s_j \ketbra{11}{00}- s^*_j\ketbra{00}{11} \;,
\end{align}
where $c_j\equiv \cos\theta_j$ and $s_j \equiv  e^{i\phi_j}\sin\theta_j$.  
This unitary generates superposition
between $\ket {00}$ and $\ket {11}$ state in the subspace of $q_0$ and $q_1$,
and acts as identity to the orthogonal subspace.
Unitary $U_{0,1}$ acting on the initial state yields
\begin{align}
U_{0,1}(\theta_1,\phi_1) \ket{00} = c_1 \ket{00} + s_1 \ket{11}\;.
\end{align}
This unitary can be realized by a circuit
as $U(\theta)$ in Figure.~\ref{fig:w_circuit}.
The state for the whole system is now
\begin{align}
c_1 \ket{000\cdots 0} + s_1 \ket{110\cdots 0}
\end{align}
Further applying $U_{0,2}(\theta_2,\phi_2)$ on $q_0$ and $q_2$ yields 
\begin{align}
c_1 c_2 \ket{000\cdots 0} 
+ c_1 s_2 \ket{101\cdots 0} 
+ s_1 \ket{110\cdots 0}
\end{align}
In this fashion we apply $U_{0,k}$ sequentially on the initial state for $k\in [1,n]$.
each application introduces amplitudes in $\ket 0 \otimes X_k {\ket 0}^{\otimes n}$ 
where the register qubit state corresponds to Hamming-one bit string with the one on the $k$-th qubit.
In order for all Hamming-weight-1 register state to be of same amplitude, and decoupled from the ancilla, the following conditions are imposed
\begin{align}
c_1c_2\cdots c_n = 0 \\
c_j |s_{j+1}| =|s_j|\;,
\end{align}
which has a solution $\sin{\theta_j} = \frac{1}{\sqrt {n+1-j}}$.
By using such angles, the $\ket{0}\otimes \ket{0}^{\otimes n}$ state is removed in the $n$-th step because $\cos \theta_n = 0$,
resulting in the final tensor-product state
\begin{align}
\ket{1} \otimes \frac{1}{\sqrt n}\big( 
e^{i\phi_1}\ket{10\cdots 0}
+e^{i\phi_2}\ket{010\cdots 0}
+\cdots
+e^{i\phi_n}\ket{0\cdots 01}
\big)
\;.
\end{align}
Setting all $\phi_j=0$ further leads to the exact $W$ state on the register:
\begin{align}
\ket{1} \otimes \ket W_n\;.
\end{align}
Figure.~\ref{fig:w_circuit} shows the corresponding circuit.

\subsection{Reverse engineering for \texorpdfstring{$W$}{W}-state preparation}
Wang, Ashhab, and Nori~\cite{wang2009efficient} outline
a procedure to produce arbitrary states with fixed particle number.
The approach is a recursive approach and scales as $O(2^{m}n^{m}/m!)$ where $m$ is the number of spin up or $\vert 1\rangle$ and $n$ is the number of qubits.
For the $W$-state case, $m=1$, the number of CNOT gates scales as $O(2n)$.

A target state that we want to prepare is transformed to the $\vert 0 \rangle^{\otimes n}$ by using a series of generalized Hadamard gates and controlled generalized Hadamard gates.  
The generalized Hadamard has the form
$\tilde{H} = C^{\dagger}XC $,
where 
\begin{align}
    C = \begin{pmatrix}u & -v^{*}\\v & u^{*}\end{pmatrix}
\end{align}
is a unitary matrix.
For any single-qubit state $\cos\theta \ket 0 + \sin\theta e^{i\phi}\ket 1$, 
it is easy to determine a $\tilde H$ that takes it to $\ket 0$:
\begin{align}
\tilde{H} = 
\begin{pmatrix}
u^{*}v +uv^{*}& (u^{*})^{2} - (v^{*})^{2}\\
u^{2} - v^{2} & - (u^{*}v +uv^{*})
\end{pmatrix}
\begin{pmatrix}
\cos\theta \\
\sin\theta e^{i\phi} \\
\end{pmatrix}
=
\begin{pmatrix}
1\\
0\\
\end{pmatrix}
\end{align}
Consider the set of states in the computational basis that 
correspond to Hamming-weight-1 bit strings, $\{X_i {\ket 0}^{\otimes n}\}_{i=1}^n$,
$W$-state is a uniform superposition of these states
\begin{align}
\ket {W_n} = \frac{1}{\sqrt{n}}\sum_{i = 1}^{n} X_i {\ket 0} ^{\otimes n}\;,
\end{align}
where $n$ indicates the number of qubits.
This wavefunction can be expressed as the first qubit and the rest $n-1$ qubits:
\begin{align}\label{eq:step1_recursion}
\vert W_n \rangle = c_{0}\vert 0 \rangle \vert W_{n-1}\rangle + c_{1}\vert 1 \rangle \vert 0\rangle^{\otimes(n-1)}
\end{align}
where $c_1=1/\sqrt{n}$ and $c_0=\sqrt{(n-1)/n}$.
Define an operator $Q_{n-1}$ that takes $\ket{W_{n-1}}$ to $\vert 0 \rangle^{\otimes (n-1)}$.
\begin{align}
Q_{n-1} \vert W_{n-1} \rangle  = \vert 0 \rangle^{\otimes (n-1)}
\end{align}
An $X$ operation on the first qubit in Eq.~\eqref{eq:step1_recursion} 
followed by a controlled-$Q_{n-1}$ operation (the first qubit being the controlling qubit)
creates the state 
$\big(c_0 \ket 0 + c_1 \ket 1 \big){\ket{0}}^{(n-1)}$,
to which we can perform the generalized Hadamard 
to evolve to the zero state $\vert 0\rangle^{\otimes n}$.  
Analogously We can define $Q_{n-1}$ using controlled-$Q_{n-2}$ and so on
until the base case for the recursion, $\ket {W_2}$, which is also the Bell state,
\begin{align}
\vert \beta \rangle = a \vert 1 0 \rangle + b \vert 0 1 \rangle
\end{align}
which can be brought to $\ket {00}$ by an $X$ gate followed by a generalized Hadamard. 
Inverse the whole circuit above gives the circuit to prepare $W$-state from the ${\ket 0}^{\otimes n}$ state.

Reference~\cite{wang2009efficient} outlines a procedure to produce arbitrary states with fixed particle number through a recursive approach that scales as  $O(2^{m}n^{m}/m!)$ where $m$ is the number of spin up or $\vert 1\rangle$ and $n$ is the number of qubits.  When considering creation of single-excitation states, or $W$-states, the circuits have $\mathcal{O}(n)$ CNOT gates single qubit gates with only nearest-neighbor physical coupling.  

 As an example for constructing an even superposition of three-excitations, a $W$-state, we provide the Quil~\cite{smith2016practical} and Cirq code below.
\begin{itemize}
\item{\bf Quil code for the W-state}
\begin{verbatim}
RY(acos(-1/3)) 2
PHASE(-pi/2) 2
RY(pi/4) 1
CNOT 2 1
RY(-pi/4) 1
RZ(pi/2) 1
CNOT 2 1
RZ(pi/2) 1
CNOT 1 0
CNOT 2 1
X 2
\end{verbatim}
\item{\bf Cirq code for the W-state}
\begin{verbatim}
qubits = cirq.LineQubit.range(3)
w_state = cirq.Circuit().from_ops([
 cirq.Ry(numpy.arccos(-1/3)).on(qubits[2]),
 cirq.ZPowGate(exponent=-1/2).on(qubits[2]),
 cirq.Ry(pi/4).on(qubits[1]),
 cirq.CNOT(qubits[2], qubits[1]),
 cirq.Ry(-pi/4).on(qubits[1]),
 cirq.Rz(pi/2).on(qubits[1]),
 cirq.CNOT(qubits[2], qubits[1]),
 cirq.Rz(pi/2).on(qubits[1]),
 cirq.CNOT(qubits[1], qubits[0]),
 cirq.CNOT(qubits[2], qubits[1]),
 cirq.X.on(qubits[2])])
\end{verbatim}
\end{itemize}

\subsection{Prepare generalized W-State via projective measurements}

This section is about preparing generalized W-states through projective measurements proposed in Ref.~\cite{Childs2002}

\paragraph{Procedure outline}
We start with the $n$-qubit state ${\ket 0}^{\otimes n}$ and apply the biased Hadamard gate
\bqa
   H=
  \left( {\begin{array}{cc}
\sqrt{1-\frac{w}{n}} & \sqrt{\frac{w}{n}} \\
   \sqrt{\frac{w}{n}} & -\sqrt{1-\frac{w}{n}} \\
  \end{array} } 
  \right)^{\otimes n}\;.
\eqa
The biased Hadamard gate will drive the initial state to $\ket \psi = H \ket 0^{\otimes n} = \big(\sqrt{1-p}\ket 0 + \sqrt{p} \ket 1\big)^{\otimes n}$ with $p=w/n$. 
Consider $\ket \psi$ measured in the computational basis, since the probability of each qubit being in $\ket 0$ is $p$, the probability of getting a state of Hamming weight $w$ is 
$\Pr (w) = C^w_n p^w (1-p)^{n-w}$,
which as a function of $w$ has a minimum at $w=n/2$ with $\Pr(w=n/2)=\frac{n!}{2^n (n/2)!(n/2)!}\approx \sqrt\frac{2}{n\pi}$, where the $\approx$ refers to large $n$ limit . Therefore $\Pr(w)\gtrsim \sqrt\frac{2}{n\pi}$ for any $w$.
Specifically, for our interest of $w=1$, we have $p=1/n$ and
\bqa
\label{eq:probw1}
\Pr(w=1)= (1-\frac{1}{n})^{n-1}=(1+\frac{1}{n-1})^{-(n-1)} \approx \frac{1}{e}\;,
\eqa
which is a fairly high probability.

If we can conduct a projective measurement on the Hamming weight (instead of measuring $\z$ on each qubit), then instead of collapsing to a state in the computational basis, the system is projected to the subspace of Hamming weight $w$.  Given Eq.~\eqref{eq:probw1}, with only a few repetitions one is expected to get $w=1$, accompanying a generalized W-state.

Now we describe the circuit to perform the projective measurement on the Hamming weight illustrated in Ref.[\onlinecite{Chuang2000}].  By definition the Hamming weight is the number of 1's of a state in the computational basis.   Computing the Hamming weight can be done by introducing an auxiliary register $W$, and apply a ``controlled-add-1'' gate on it.  The ``add-1" is conditioned on the qubit being in state 1.  
Since the Hamming weight is at most $n$, $\log n$ qubits are sufficient to encode $W$.  Upon measuring the auxiliary qubits, one gets a generalized W-state when the measured $W=1$. Now we consider how the controlled-add-one is implemented.
First, we review a more general operation controlled-add-k for any constant $k$ presented in Ref.[\onlinecite{Cleve1996}].
The idea is to use a $n$-bit auxiliary register $C$ to record whether a carry will happen for next bit.  Populating $C$ bit by bit.  In a second loop, do the real addition.

Program {\tt\bf Conditional\_Add\_k}\\
\begin{itemize}
\item Notation\\
X : n-bit register \\
B : bit register (control) \\
C : n-bit auxiliary register (initialized and finalized to 0)\\
\item Pseudo Code
\end{itemize}
\begin{empheq}[box=\widefbox]{align}
& {\tt\bf Conditional\_Add\_k:}\nonumber\\
  &\text{for $i = 1$ up to $n-1$} \\
  \label{eq:assignC}
  &~~~~ C_i \leftarrow C_i \oplus \text{MAJ}(k_{i-1},X_{i-1},C_{i-1}) \\
  &\text{endfor} \nonumber\\
  &\text{for $i = n-1$ down to 1} \\
  \label{eq:addk}
  &~~~~ X_i \leftarrow X_i\oplus (k_i \land B) \\
  \label{eq:addC}
  &~~~~ X_i \leftarrow X_i\oplus (C_i \land B) \\
    \label{eq:resetC}
  &~~~~ C_i \leftarrow C_i \oplus \text{MAJ} (k_{i-1}, X_{i-1}, C_{i-1}) \\
  &\text{endfor} \nonumber\\
\label{eq:flipX0}
  &X_0 \leftarrow X_0 \oplus k_0
\end{empheq}

where the ``majority" gate MAJ takes value true when at least 2 out of the three bits are true:

\begin{eqnarray}
\text{MAJ}(O,P,Q) =
\begin{cases}
P\land Q & \text{if $O=0$} \\
P\lor Q & \text{if $O=1$} 
\end{cases}
\end{eqnarray}

We now explain the above pseudo-code.\\
Line~\ref{eq:assignC}, assuming $B=1$, i.e., the addition will happen, determines whether a carry will happen for the next bit, and record in $C_i$.
A Toffli (controlled-controlled-not) gate is applied to $C_i$, with the control condition being that at least two bits in $\{k_{i-1},X_{i-1},C_{i-1}\}$ take value one. 
Because $C$ is initialized to be all zeros, the control condition being true will set $C_i=1$. \\
Line~\ref{eq:addk} and Line~\ref{eq:addC} implements the real addition of $k_i$ and $C_i$ to $X_i$, controlled by $B$.\\
Line~\ref{eq:resetC} is exactly the same as Line~\ref{eq:assignC}, hence resets $C$ to zero.\\
Line~\ref{eq:flipX0} adds $k_0$ to $X_0$.

Now we consider the special case $k=1$.  We only need a single-bit register for $k$: $k_0=1$. 
Line~\eqref{eq:assignC} reduce to 
$C_1=C_1 \oplus (X_0 \lor C_0)$
and
$
C_i=C_i \oplus (X_{i-1} \land C_{i-1})
$
for $i>1$.
Therefore the pseudo-code is

Pseudo-code:\\
\begin{empheq}[box=\widefbox]{align}
& {\tt\bf Conditional\_Add\_1:}\nonumber\\
  &\text{for $i = 1$ up to $n-1$} \\
  \label{eq:assignC1}
    &~~~~ \tt{SET\_C}\\
  &\text{endfor} \nonumber\\
  &\text{for $i = n-1$ down to 1} \\
  \label{eq:addC1}
  &~~~~ X_i \leftarrow X_i\oplus (C_i \land B) \\
    &~~~~ \tt{SET\_C}\\
    &\text{endfor} \nonumber\\
\label{eq:flipX01}
  &X_0 \leftarrow \bar X_0
\end{empheq}
 
with subrutine 
 \begin{empheq}[box=\widefbox]{align}
    &~~~~ \tt{SET\_C:}\\
    &~~~~ \text{if $i==1$}\\
  &~~~~ ~~~~ C_i \leftarrow C_i \oplus (X_{i-1}\lor C_{i-1}) \\
    &~~~~\text{else}\nonumber\\
    &~~~~~~~~ C_i \leftarrow C_i \oplus (X_{i-1}\land C_{i-1}) \\
      &~~~~\text{endif}
\end{empheq}

Number of gates: The circuit for computing the Hamming weight would involve $n$  {\tt Controlled\_add\_1} gates on the Hamming weight register $W$, each controlled by one data qubit.
Since $W$ is composed of $\log n$ bits, each {\tt Controlled\_add\_1} requires $5\log n$ Toffoli (or Toffoli-like) gates. 
The total circuit requires $5n\log n$ Toffoli gates.
With the probability $1/e$ of getting $W=1$ in the measurement, it will take on average three repetition, hence the expected number of Toffoli gates is $\sim 15 n \log n$, which translates into $\sim 90 n \log n$ CNot gates.

Note that when measurement is expensive, it would be more efficient to hold the measurement on $W$ after performing the computation of $W$, apply the following target algorithm, and measure $W$ in the end.  This will triple the run time of the whole algorithm.

\subsection{Applying \texorpdfstring{$XY$}{XY} Hamiltonian}
\label{sec:XY_W}
The $XY$ Hamiltonian on a 1D chain or ring
can be exactly implemented in gate-model.
Applying the $XY$ Hamiltonian on a state $\ket 0$ can generate certain superposition of states in the subspace $S$ expanded
by states corresponding to Hamming-weight-1 bit strings, 
but exact W-state may not be generated this way.
The Hamiltonian for a 1D $XY$ model with n.n. coupling reads
\bqa
H_{XY} \eq \sum_{c=1}^m \x_{c}\x_{c+1}+\y_{c}\y_{c+1} \\
\eq \frac{1}{2}  \sum_{c=1}^m \big(\sigma^+_{c}\sigma^-_{c+1}+\sigma^-_{c}\sigma^+_{c+1}\big)
\eqa
where periodic boundary condition (PBC) $\sigma_{m+1}=\sigma_{1}$ is implied.
We now examine how well the ring transfers the a classical Hammint-weight-1 state.

\subsubsection{State transfer using a \texorpdfstring{$XY$}{XY} chain with open boundary condition (OBC)}
If we remove the constraint of PBC, i.e., work on a open-end chain instead of ring, it is known that perfect state transfer can be achieved along a $XY$ chain only for $n=2$ and $n=3$.
The fidelity can be computed by diagonalizing the Hamiltonian, in this case the eigenvectors reads
\bqa
\ket k = \sqrt{\frac{2}{m+1}} \sum_{n=1}^m \sin(\frac{nk\pi}{N+1}) \ket n
\eqa
and the eigenvalues are $E_k=2\cos(\frac{k\pi}{m+1})$.
Furthermore, if inhomogeneous coupling between n.n. qubits along the chain is allowed, perfect transfer can be realized for any chain length.\cite{Christandl04}

\subsubsection{State transfer using a \texorpdfstring{$XY$}{XY} chain with PBC}
We now study how the state $\ket 0$ is transferred along a ring. 
We apply Jordan-Wigner transformation~\cite{lieb_two_1961, barouch_statistical_1970} to $H_{B,j}$. 

\bqa
 a_c \eq  S^-_c e^{i\phi_c} \\
 a_c^\dag \eq  S^+_c e^{-i\phi_c}
\eqa
where $S^+_c=(\x_c+i\y_c)/2$, $S_c^-=(\x_c-i\y_c)/2$, and the phase factor $\phi_c=\pi \sum_{c'<j} (\z_{c'}+1)/2$ is long-ranged involving all operators for $c'<c$.
The new operators $a_c,~a_c^\dag$ can be verified to obey the fermion anticommutation relations, 
$\{a_c,a_{c'}^\dag\}=a_ca_{c'}^\dag+a_{c'}^\dag a_c=\delta_{c,c'}$, and
$\{a_c,a_{c'}\}=\{a_{c}^\dag,a_{c'}^\dag\}=0$.
The inverse transformation reads
\bqa
S_c^+ \eq  a_c^\dag e^{i\phi_c}\\
S_c^- \eq  a_c e^{-i\phi_c}\\
\z_c \eq 2 a^\dag_c a_c-1 
\eqa
and the phase factor in the fermionic representation is $\phi_c = \pi \sum_{c'<c} a^\dag_{c'} a_{c'}$.
The Jordan-Wigner transformation is a convenient tool for one-dimentional spin systems,
particularly for nearest-neighbored couplings because in product of the neighboring spin operators like $S_c^+S_{c+1}^-$, the phase factors drop out, leaving a concise expression with short-ranged coupling.

Apply the transformation to our problem, For simplicity, we omit the index $j$ in the Pauli operators for this section, and without loss of generality we use ${\cal P}_j=\{1,2,\ldots,m\}$ . We get
\bqa
H_{XY} 
\eq 2 \sum_{c=1}^{m-1} a_c^\dag a_{c+1} - a^\dag_m a_1 G + h.c.  
\label{eq:HBj_fermion}
\eqa
and the initial state $\ket{1,0,\cdots,0}$ in the fermionic representation becomes 
\bqa 
\ket \Phi_0=a^\dag_1 \ket{\bf 0}
\eqa 
where $\ket{\bf 0}$ denotes the vacuum state. (the 0-eigenstate of the total particle number operator $\hat N_\text{tot} = \sum_c a^\dag_c a_c$).
Here we introduced gauge operator $G=\exp[i\pi \sum_{l=1}^m a^\dag_l a_l]=\prod_{l=1}^m (-) \z_l$.  The initial state has only one spin up, hence corresponds to $G=-1$.  Because $G$ commutes with both $H_B$ and $H_C$, it is a constant of motion and its value statys $-1$ throughout the evolution.  We can therefore replace $G$ with $-1$ in Eq.~\eqref{eq:HBj_fermion}:
\bqa
H_{XY} 
\eq 2 \sum_{c=1}^{m} a_c^\dag a_{c+1} + h.c.
\eqa
This Hamiltonian can be diagonalized by introducing the Fourier transformation $f_k=\frac{1}{\sqrt m}\sum_{c=1}^m \exp[-ick\omega]a_c$, with $\omega=\frac{2\pi}{N}$. 
The diagonalized Hamiltonian is
\bqa
H_{XY} \eq \sum_{k=1}^{m} E_k f_k^\dag f_k  \nnm
\eqa
where the eigen-energies $E_k=2\cos(2k\pi/m)$
and the eigen states are 
\bqa
\ket {\psi_k} \eq f^\dag_k\ket 0 
= \frac{1}{\sqrt m} \sum_{c=1}^m e^{ick\omega} a^\dag_c \ket 0 
\eqa

We now measure the expectation value of the occupation operator $\hat n_c=a^\dag_c a_c$ for each site $c$.
The occupation operator $\hat n_c$ taking value 1 indicates the the spinless fermion particle is at site $c$, and correspondingly in the spin problem the spin on site $c$ is up.

At time $t$ the fidelity of sate transfer at site $c$ can be analytically derived to be
\bqa
{\cal F} &\equiv& \Tr[\hat n(c) e^{-itH_{XY}}\ketbra{\Phi_0}{\Phi_0} e^{itH_{XY}} ] \\
\eq \Big| \frac{1}{m} \sum_k   e^{i(c-1)k\omega} e^{-itE_k} \Big|^2
\eqa

\begin{figure}[htbp]
	\begin{center}  
		\includegraphics[width=0.5\textwidth]{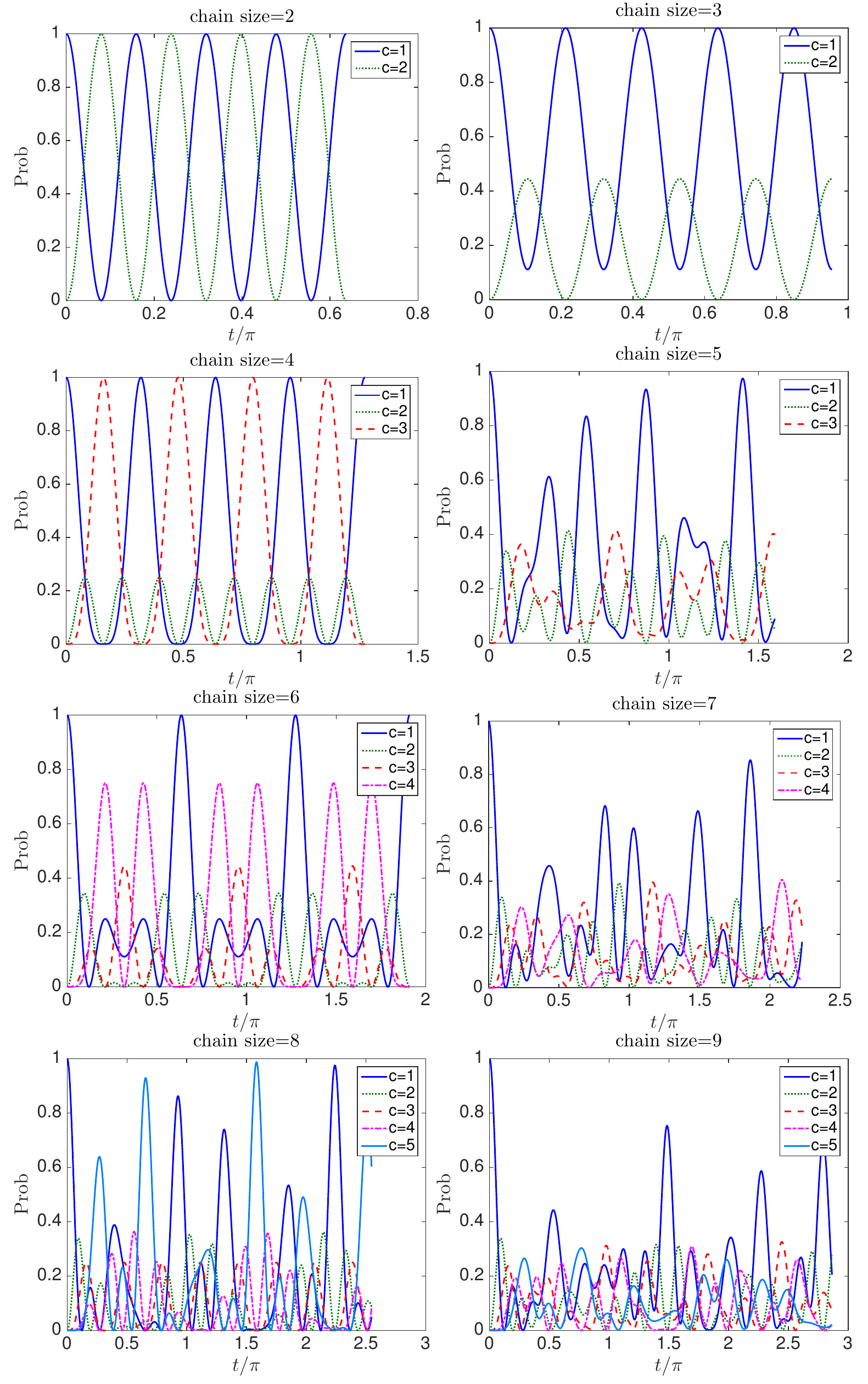}
		\caption{
			For ring size $m=2$ to 9, probability of spin-up at each site as a function of time $t$, under the evolution of the XY Hamiltonian on the ring.
			\label{fig:probRing}
			}  
	\end{center} 
\end{figure}

Figure.~\ref{fig:probRing} shows for the problability evolution of the transfer fidelity at each site. 
For W-state generation, we are looking for a time when pupulation on all sites has equal probability, i.e., when curves of different color interacts at $1/n$.  We see for $n$ up to 4, an exact W-state can be generated by choosing the right time $t$.

\bibliography{biblo}
\end{document}